\documentclass[prd,onecolumn,notitlepage,
nofootinbib,aps,tightenlines,
preprintnumbers,amsmath,amssymb,amsfonts,showpacs,superscriptaddress]{revtex4-1}

\usepackage[hyperindex,breaklinks]{hyperref}
\usepackage{graphicx,epstopdf}

\usepackage{natbib,bm}
\usepackage[usenames,dvipsnames]{color}
\usepackage{slashed}    
\usepackage{hyperref}
\usepackage{breakurl}
\usepackage{multirow}
\usepackage{bbold}
\usepackage{listings}
\usepackage{comment}

\usepackage{mathrsfs}
\newcommand{\order}[1]{\mathcal{O}{(#1)}}
\newcommand{\m}{\text{m}}
\newcommand{\cm}{\text{cm}}
\newcommand{\GeV}{\text{GeV}}
\newcommand{\eV}{\text{eV}}
\newcommand{\MeV}{\text{MeV}}
\newcommand{\x}{\chi}

\newcommand{\mxx}{m_{\chi \chi}}
\newcommand{\Ap}{A^\prime}
\newcommand{\mAp}{m_{A^\prime}}
\newcommand{\Lag}{\mathscr{L}}
\newcommand{\eps}{\epsilon}
\newcommand{\Nova}{NO$\nu$A}

\def\be{\begin{equation}}
\def\ee{\end{equation}}
\def\ba{\begin{eqnarray}}
\def\ea{\end{eqnarray}}
\def\ge{\mathrel{\raise.3ex\hbox{$>$\kern-.75em\lower1ex\hbox{$\sim$}}}}
\def\la{\mathrel{\raise.3ex\hbox{$<$\kern-.75em\lower1ex\hbox{$\sim$}}}}

\def\simgt{\mathrel{\raise.3ex\hbox{$>$\kern-.75em\lower1ex\hbox{$\sim$}}}}
\def\simlt{\mathrel{\raise.3ex\hbox{$<$\kern-.75em\lower1ex\hbox{$\sim$}}}}

\newcommand{\nc}{\newcommand}
\nc{\al}{\alpha}
\nc{\ga}{\gamma}
\nc{\de}{\delta}
\nc{\ep}{\epsilon}
\nc{\ze}{\zeta}
\nc{\et}{\eta}
\nc{\ka}{\kappa}
\nc{\rh}{\rho}
\nc{\si}{\sigma}
\nc{\ta}{\tau}
\nc{\up}{\upsilon}
\nc{\ph}{\phi}
\nc{\ch}{\chi}
\nc{\ps}{\psi}
\nc{\om}{\omega}
\nc{\Ga}{\Gamma}
\nc{\De}{\Delta}
\nc{\La}{\Lambda}
\nc{\Si}{\Sigma}
\nc{\Up}{\Upsilon}
\nc{\Ph}{\Phi}
\nc{\Ps}{\Psi}
\nc{\Om}{\Omega}
\nc{\ptl}{\partial}
\nc{\del}{\nabla}
\nc{\ov}{\overline}
\nc{\newcaption}[1]{\centerline{\parbox{15cm}{\caption{#1}}}}
\nc{\us}{U(1)$_S$}

\mathchardef\mhyphen="2D

\def\beq{\begin{equation}}
\def\eeq{\end{equation}}
\def\bmat{\begin{displaymath}}
\def\emat{\end{displaymath}}
\def\bear{\begin{eqnarray}}
\def\eear{\end{eqnarray}}
\def\ba{\begin{eqnarray}}
\def\ea{\end{eqnarray}}
\def\bery{\begin{array}}
\def\ery{\end{array}}
\def\bit{\begin{itemize}}
\def\eit{\end{itemize}}
\def\ben{\begin{enumerate}}
\def\een{\end{enumerate}}
\def\btab{\begin{tabular}}
\def\etab{\end{tabular}}
\def\btbl{\begin{table}}
\def\etbl{\end{table}}
\def\bfig{\begin{figure}[htb]}
\def\efig{\end{figure}}
\def\bpic{\begin{picture}}
\def\epic{\end{picture}}

\def\ga{\mathrel{\raise.3ex\hbox{$>$\kern-.75em\lower1ex\hbox{$\sim$}}}}
\def\la{\mathrel{\raise.3ex\hbox{$<$\kern-.75em\lower1ex\hbox{$\sim$}}}}
\def\gappeq{\mathrel{\rlap {\raise.5ex\hbox{$>$}}
{\lower.5ex\hbox{$\sim$}}}}
\def\lappeq{\mathrel{\rlap{\raise.5ex\hbox{$<$}}
{\lower.5ex\hbox{$\sim$}}}}

\def\gyr{{\rm \, G\kern-0.125em yr}}
\def\mev{{\rm \, Me\kern-0.125em V}}
\def\gev{{\rm \, Ge\kern-0.125em V}}
\def\tev{{\rm \, Te\kern-0.125em V}}

\newcommand{\nla}{\nonumber \\}

\begin{document}

\title{On sub-GeV Dark Matter Production at Fixed-Target Experiments}

\author{Asher Berlin}
\affiliation{Center for Cosmology and Particle Physics, Department of Physics, New York University, New York, NY, USA}

\author{Patrick deNiverville}
\affiliation{Center for Theoretical Physics of the Universe, IBS, Daejeon 34126, Korea}
\affiliation{T2, Los Alamos National Laboratory (LANL), Los Alamos, NM, USA}

\author{Adam Ritz}
\affiliation{Department of Physics and Astronomy, University of Victoria, 
Victoria, BC V8P 5C2, Canada}

\author{Philip Schuster}
\affiliation{SLAC National Accelerator Laboratory, 2575 Sand Hill Road, Menlo Park, CA, USA}

\author{Natalia Toro}
\affiliation{SLAC National Accelerator Laboratory, 2575 Sand Hill Road, Menlo Park, CA, USA}

\date{\today}

\begin{abstract}

We analyze the sensitivity of fixed-target experiments to sub-GeV thermal relic dark matter models, accounting for variations in both mediator and dark matter mass, and including dark matter production through both on- and off-shell mediators.  It is commonly thought that the sensitivity of such experiments is predicated on the existence of an on-shell mediator that is produced and then decays to dark matter.  While accelerators do provide a unique opportunity to probe the mediator directly, our analysis demonstrates that their sensitivity extends beyond this commonly discussed regime.  In particular, we provide sensitivity calculations that extend into both the effective field theory regime where the mediator is much heavier than the dark matter and the regime of an off-shell mediator lighter than a dark matter particle-antiparticle pair. 
Our calculations also elucidate the resonance regime, making it clear that all but a fine-tuned region of thermal freeze-out parameter space for a range of simple models is well covered. 

\end{abstract}
\maketitle


\section{Introduction}
\label{sec:intro}

The evidence for a dark matter (DM) component in the universe, through its gravitational effects over many distance scales, provides the strongest indicator to-date for physics beyond the Standard Model (SM). Over the past decade, theoretical and experimental approaches to particle DM have evolved significantly, with a loosening of theoretical priors about how DM may be explained within particle physics. This has been driven on the one hand by the increasingly stringent LHC constraints on new TeV-scale degrees of freedom, and on the other by the recognition that the strong empirical evidence for DM motivates exploring all viable and testable scenarios for DM, not just those that are linked to other expectations for new physics (e.g., the naturalness problem motivating electroweak-scale WIMPs, or the strong CP problem motivating axions).

Within this broader perspective, there is strong theoretical motivation for considering \emph{thermal relic} models where the DM was once in thermal equilibrium with ordinary matter. This assumption constitutes a ``minimal''  cosmology in two important ways: the cosmological evolution of DM has minimal dependence on initial cosmological conditions, and at the same time closely mimics the early-universe production processes known to accurately describe Big Bang nucleosynthesis (BBN).  
The full mass range for such thermal relics is wider than for conventional electroweak WIMPs, which must be heavier than a few GeV as a result of the Lee-Weinberg bound~\cite{LW}. The {\it light DM} window from an MeV to a GeV is viable for DM that is neutral under SM gauge groups but possesses distinct interactions~\cite{Boehm:2003hm}. The lower boundary of the light DM mass range is motivated by the successful predictions of BBN and observations of the cosmic microwave background (CMB) (see Refs.~\cite{Berlin:2017ftj,Berlin:2018ztp} for an exception).  The upper boundary is purely conventional, as heavier hidden-sector DM has increasingly WIMP-like phenomenology. 
The light DM paradigm has been explored in some depth over the past 5-10 years~\cite{DS16,CV17,PBC} in both direct detection~\cite{Essig:2011nj,Essig:2012yx,Angloher:2015ewa,Hochberg:2015pha,Essig:2015cda,Abdelhameed:2019hmk,Abramoff:2019dfb} and accelerator-based experiments using electron~\cite{Bjorken:2009mm,Izaguirre:2013uxa,Diamond:2013oda,Izaguirre:2014dua,Batell:2014mga,Lees:2017lec,Berlin:2018bsc,NA64:2019imj} and proton~\cite{Batell:2009di,deNiverville:2011it,deNiverville:2012ij,Kahn:2014sra,Adams:2013qkq,Soper:2014ska,Dobrescu:2014ita,Coloma:2015pih,dNCPR,MB1,MB2,Alpigiani:2018fgd,Ariga:2018pin} beams (see also Refs.~\cite{Hewett:2012ns,Kronfeld:2013uoa,Essig:2013lka,pospelov2008,Batell:2009yf,Essig:2009nc,Reece:2009un,Bjorken:2009mm,Freytsis:2009bh,Batell:2009jf,Freytsis:2009ct,Essig:2010xa,Essig:2010gu,McDonald:2010fe,Williams:2011qb,Abrahamyan:2011gv,Archilli:2011zc,Lees:2012ra,Davoudiasl:2012ag,Kahn:2012br,Andreas:2012mt,Essig:2013vha,Davoudiasl:2013jma,Morrissey:2014yma,Babusci:2014sta,Izaguirre:2015yja} for studies of related dark sectors). Yet the multi-dimensional parameter space of light DM models poses a challenge for assessing both the status of the field and prospects  for upcoming searches.

From a general effective field theory (EFT) perspective, the leading couplings of the SM to a fully neutral dark sector could arise from vector, Higgs, or neutrino portals, which rely on the presence of an associated vector ($A'_\mu$), scalar ($S$), or singlet fermion ($N$) mediator, respectively. These portals are the unique marginal or relevant operators that can couple the two sectors, and the neutrino portal provides the simplest model of neutrino mass.
Thermal freeze-out is generically characterized by the requirement of a weak scale annihilation cross section, $\langle \si_{\rm ann} v \rangle \sim 1 \ \text{pb}$. This ensures that the frozen out mass density matches the observed relic abundance of DM. For orientation, we can parametrize the interactions between DM fields, $\ch$, and the SM through a set of higher-dimensional operators of the form
\begin{align}
 \Lag_{\rm EFT} &= \frac{1}{\La^n} \, {\cal J}^{\mu\ldots}_D \, {\cal J}^{\rm SM}_{\mu\ldots} +\cdots, \label{EFT}
\end{align}
where ${\cal J}^{\mu\ldots}$ denotes a set of (pseudo-)scalar, (axial-)vector, or other currents associated with the mediation channels between DM (${\cal J}_D$) and SM fields (${\cal J}_{\rm SM}$).  
In the minimal case that $2\rightarrow2$ processes mediate DM annihilation through dimension-six interactions, we have $n=2$ in Eq.~\eqref{EFT} and thermal freeze-out implies the parametric relation 
\be
 \langle \si_{\rm ann} v \rangle \sim c ~ \frac{m_\ch^2}{2\pi\La^4} \sim 1 \ {\rm pb} \label{tt}
 \, ,
\ee
where $c$ is a numerical factor that depends on spins and other quantum numbers of the currents. This scaling explains why WIMPs (with $m_\chi \sim \La\sim \text{TeV}$) are archetypal DM candidates, but allows a broad range of DM masses when $\La$ is allowed to vary. The relation of Eq.~(\ref{tt}) is a schematic form of the {\it thermal target} that models must satisfy in order to reproduce all of the observed DM and must at a minimum exceed in order not to overpopulate the matter content of the universe.  

An operator description, as in Eq.~\eqref{EFT}, is valid at energy scales below the mass of the force mediators responsible for coupling SM particles to DM.  If the force mediator mass scale is experimentally accessible or relevant during thermal freeze-out, then an accurate description of DM phenomenology and/or cosmology requires a more complete theory --- one that includes the mediator particle explicitly, rather than integrating it out as in Eq.~\eqref{EFT}.  Indeed, the presence of light dark force mediators is a critical ingredient to bypassing the Lee-Weinberg bound.  Thus, light DM must form part of a multi-component dark sector, and it is reasonable to expect that the mediator mass may be relatively near the DM mass.  

Motivated by models of light DM, much theoretical and experimental effort over the past decade has focused on the vector portal, which at low energies involves kinetic mixing between the photon and the dark vector,  $(\ep/2) \, F^{\mu\nu}F'_{\mu\nu}$. This is in part because it is the least constrained scenario that allows for bilinear mixing. It also provides the most scope for model building, including what has become the benchmark model for sub-GeV DM~\cite{DS16,CV17}.  We will consider the vector portal, and modest generalizations of it, as our primary examples below as we discuss the domain of validity of effective theory and the full parameter space of DM detection.  

Under the assumptions of Eq.~\eqref{tt}, the EFT is always valid for the kinematics of direct detection, which involves momentum transfers parametrically smaller than the DM particle mass. 
The most convenient description of the interaction is via a relativistic (or non-relativistic) EFT for DM interactions with nucleons or electrons, which is entirely determined by Eq.~\eqref{EFT}.  For example, for fermionic DM $\x$ coupled to SM vector currents, the leading interactions are typically of the form 
\begin{align}
 \Lag_{\rm EFT} &=  \frac{1}{\La_e^2} ~ \bar{\ch} \gamma^\mu \ch  ~ \bar{e}\gamma^{\mu} e + \frac{1}{\La_p^2}  ~ \bar{\ch} \gamma^\mu \ch ~ \bar{p} \gamma_\mu p + \cdots
 ~.
 \label{eft2}
\end{align}
The scattering cross section on nucleons then depends parametrically on $\mu_{N,\ch}^2/\La^4$, with $\mu_{N,\ch}$ the reduced DM-nucleon mass, and so direct detection experiments simply constrain the cutoff scale $\La$ for a given DM mass. 

\begin{figure}[t]
\centerline{
\includegraphics[width=10cm]{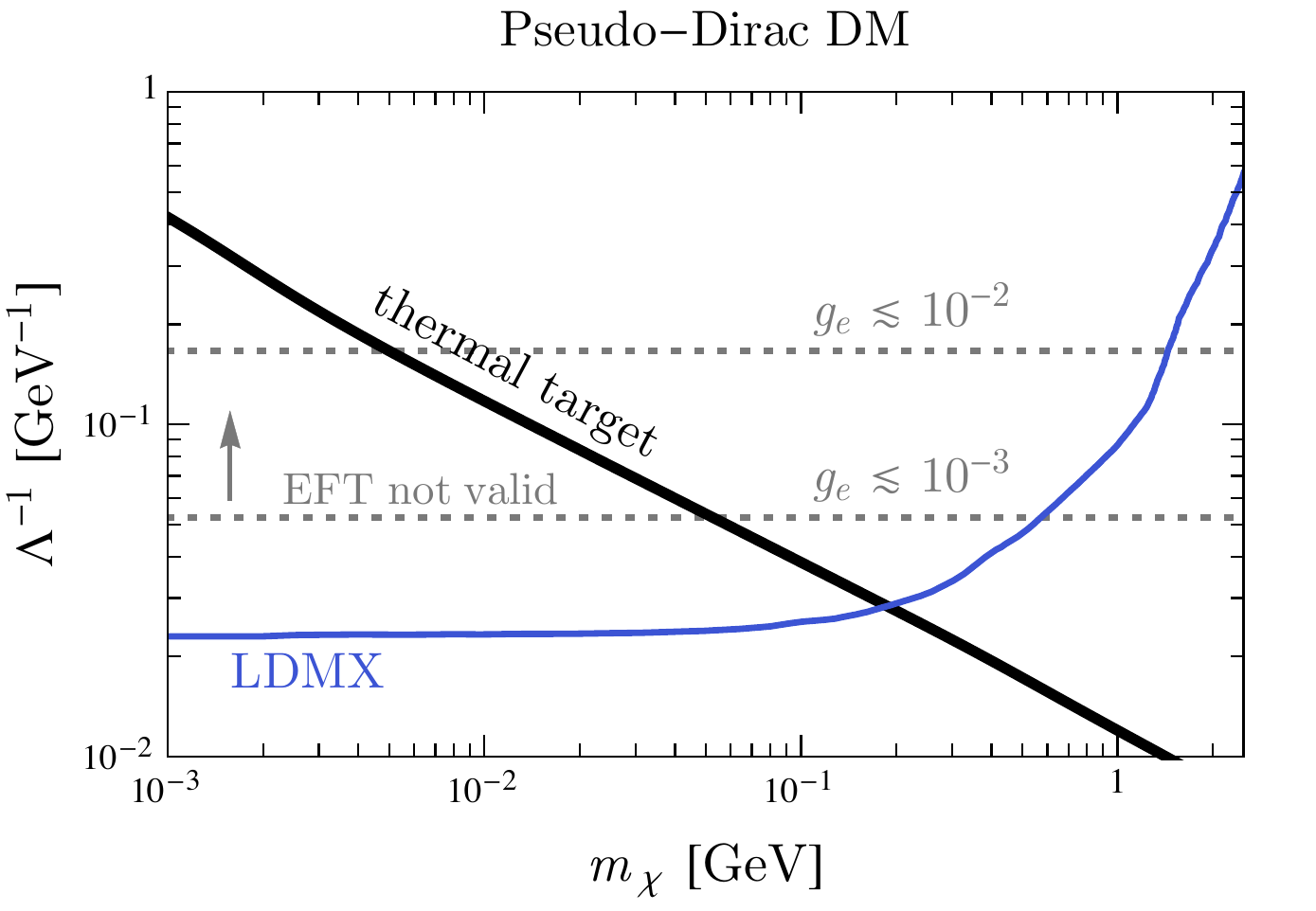}
}
\caption{For an example EFT contact operator of the form $\Lag_{\rm EFT} \supset i \, \overline{\x}_1 \gamma^\mu \x_2 ~ \overline{e} \gamma_\mu e /  \La^{2} $ (where $\x_{1,2}$ are a nearly degenerate pseudo-Dirac pair of Majorana fermions), the thermal target (black) and sensitivity at the proposed LDMX electron beam missing-momentum experiment (blue) are shown for a range of dark matter masses. The EFT is  taken to be valid if $\La \gtrsim \sqrt{s}/(g_e \, g_D)^{1/2}$, where $g_e$ is the mediator-electron coupling, $g_D \lesssim 4 \pi$, and for LDMX $s=4E_e E_\gamma$ with the typical photon energy $E_\gamma$ given by the nuclear size.  This condition ensures that $\sqrt{s} < m_V$, where $m_V$ is the mediator mass. The EFT is invalid in regions above the gray dotted line, for various choices of the mediator-electron coupling. These dotted lines move down in the figure as the SM couplings are reduced.} 
\label{figEFT}
\end{figure}

By contrast, the usual analysis of accelerator-based searches starts from the assumption that the mediator is, in fact, light enough to be produced.  We shall see that this assumption is often reasonable, but it is illustrative to instead begin from the EFT description (Eq.~\eqref{eft2}) of the DM-SM interaction.  The result of this analysis is illustrated in Fig.~\ref{figEFT} for a particular  operator (see Eq.~(\ref{eq:DiracCurrent}) below) and a particular experiment (the proposed LDMX experiment, see Sec.~\ref{sec:summary} below).  Other operators of the same dimension lead to results that are qualitatively very similar. 
For a given DM mass $m_\chi$ on the horizontal axis of Fig.~\ref{figEFT}, smaller values of $\Lambda^{-1}$ on the  vertical axis correspond to greater sensitivity.  This illustrates LDMX's sensitivity, within the EFT treatment, to the interaction strengths expected for thermal freeze-out for DM particle masses below $\sim 200$ MeV.

The maximum momentum transfer in the kinematics explored by LDMX is $\sqrt{s}\sim \text{GeV} $, such that the EFT of Eq.~\eqref{eft2} does not become strongly coupled anywhere within the range of Fig.~\ref{figEFT}. Nonetheless, there is reason to believe that the EFT has a smaller region of validity.  Concretely, other constraints at GeV energy scales motivate the assumption that the interaction of Eq.~\eqref{eft2} is mediated by tree-level exchange of a mediator (of mass $m_V$) that couples to SM fields with interaction strengths $g_{\rm SM}$ and to DM with strength $g_D$.  In this case, we expect 
\be
 \La^2 \propto \frac{m^2_V}{g_{\rm SM} \, g_D} ~ \label{cutoff}
\ee
(other high-energy completions for the contact interaction of Eq.~\eqref{eft2} would change the constant of proportionality, while maintaining $\La \propto m_V$).
Constraints on $g_{\rm SM}$ from electron self-interactions imply $g_e \lesssim \order{10^{-3}} - \order{10^{-2}}$ depending on the precise mediator mass.  
The EFT treatment is valid only for $m_V >\sqrt{s}$ --- a constraint that (using Eq.~\eqref{cutoff} and the constraints on $g_e$) is \emph{inconsistent} with large $\Lambda^{-1}$, and is also invalid in models where a small $\Lambda^{-1}$ results from a light mediator with weaker coupling.
A further complication to the EFT interpretation of accelerator-based searches is that different experiments (and even different production modes relevant within a single experiment) have different maximum energies $\sqrt{s}$, so that the regions of consistency of the EFT vary from one experiment to another.  We emphasize, however, that reaching beyond the regime where EFT is valid is not a deficit of accelerator-based experiments --- rather, it is a signal that they can potentially uncover the mediation mechanism behind the freeze-out process.  

For these reasons, the experimental constraints should instead be evaluated in the larger parameter space:
\be
 \mbox{Light DM Parameters} = \{m_{\rm V}, m_{\rm DM}, g_{\rm SM}, g_D\}.
\ee
This same enlarged parameter space is also motivated by the calculation of the DM relic abundance resulting from thermal freeze-out.  When $m_{\rm V}\gg m_{\rm DM}$, the DM abundance reduces to a function of the cutoff $\La^2 \propto \frac{m^2_V}{g_{\rm SM} \, g_D}$, but for $m_{\rm V}\lesssim 2 \, m_{\rm DM}$ 
the thermal abundance has more complex dependence on the four parameters above. 

Practical limitations motivate examining the sensitivity of different experimental probes in two-dimensional slices of this parameter space, of which several have been considered in the literature.  The most commonly used is to fix $m_{\rm V} = 3 \, m_{\rm DM}$, which generally yields a conservative estimate of experimental sensitivity within the region where the EFT is approximately valid at freeze-out and mediator production is kinematically accessible.  The resulting 3-dimensional parameter space is further reduced by either saturating $g_D$ at a large value near the perturbativity bound or fixing the product $g_{\rm SM} \cdot g_D$ to the value predicted by thermal freeze-out for given $m_{\rm DM}$.  However, fixing the ratio of $m_{\rm V}$ to $m_{\rm DM}$ obscures some important features of the parameter space and of experiments' sensitivities, particularly in the EFT limit and the resonance region (the latter was the focus of Ref.~\cite{Feng:2017drg}).

With the experimental effort to fully explore light DM now becoming a reality,\footnote{E.g., the DOE allocated dedicated funding to new light DM search experiments in 2019.} it is timely to analyze the full thermal relic parameter space, in order to determine what capabilities are needed to fully test and either discover or exclude simple models of light thermal DM. In this paper, we initiate such an analysis and uncover some generic features of these models that were less visible in the conventional parameter slices considered thus far. In addition, the extended coverage allows the sensitivity to be tracked from the EFT limit, with heavy off-shell mediators, to the resonance region where these mediators may be produced on-shell in fixed-target experiments, and all the way to off-shell light mediators below the DM pair threshold. For concreteness, our analysis will focus on models that involve a dark photon mediator, but the features uncovered are generic for models within this dark sector paradigm.

In the next section, we summarize and motivate the thermal DM framework in more detail. In Sec.~\ref{sec:Rintro}, we introduce a new parameter plane, which allows for a more comprehensive analysis of the reach of experiments. In Secs.~\ref{sec:summary} and \ref{sec:production}, we summarize the fixed-target experimental facilities to be analyzed and the primary production modes. Our results, presenting a comprehensive picture of the reach of these facilities, appears in Sec.~\ref{sec:sensitivity}, and we finish with some concluding remarks in Sec.~\ref{sec:conc}.

\section{Dark sectors and predictive light DM models}

In this section, we will briefly review the assumptions made in constructing light DM models in the thermal relic window. Fermion models of DM are especially natural to consider for several reasons. First, all stable matter that we know of is comprised of fundamental fermions -- it would be natural for this to be true of DM as well. Second, the stability of new fermions is rather easy to realize if the analog of lepton or baryon number, i.e., ``dark lepton number,'' is preserved. This is a rather natural feature of many models, especially those where the fermions are coupled to a vector field $\Ap$, as illustrated below. Third, fermion masses can be technically natural -- radiatively insensitive to unknown UV physics. 

The fermion model with the fewest number of degrees of freedom, and therefore the most ``minimal'' DM model in some sense, consists of a single two-component Weyl fermion $\xi$. In this case, the vector and scalar interactions are uniquely fixed by relativistic covariance to be, 
\be
\label{eq:DMcurrents}
{\cal  J}_D^\mu = \xi^\dagger \overline\sigma^\mu \xi = - \frac{1}{2} \,  \overline \x \gamma^\mu \gamma^5 \x 
~~,~~
{\cal J}_D = \xi^2 + \text{h.c.} = \overline{\x} \x
~, 
\ee
where we have grouped $\xi$ into a 4-component Majorana spinor $\x = (\xi~~ \xi^\dagger)^T$. Without additional structure, ``$\xi$-number" is automatically conserved, so $\x$-particles will be stable. Majorana fermion DM arises naturally in supersymmetric dark sector models and more generally in scenarios involving additional $U(1)_D$ gauge groups, such that $U(1)_D$-breaking mass terms are larger than $U(1)_D$-preserving mass terms. 

A well-motivated hypothesis for the origin of DM is that it reached thermal equilibrium with ordinary matter in the hot early universe, then ``froze out'' of equilibrium as the universe cooled, leaving a residual DM abundance surviving to the present day.  This \emph{thermal DM} scenario is both simple, readily achieving the observed DM abundance,  and predictive,  requiring a precise cross section for DM annihilation into ordinary matter to match the observed density.  In many models, this in turn implies a minimum interaction strength for DM, which serves as an important sensitivity benchmark for DM searches. 

For DM lighter than a GeV, DM couplings to the lightest SM species (such as the SM fermions, $f$) can easily explain DM's thermal origin. At low-energies, such couplings can be parametrized as four-Fermi contact operators suppressed by the mass-scale of the interaction
\be
\label{eq:DMcurrents2}
 \Lag = \frac{c_\text{v}}{\La^2} ~ {\cal J}^\mu_D  \, J^\mu_f + \frac{c_\text{s}}{\La^2} ~ {\cal J}_D  \, J_f
 ~,
\ee
where $c_\text{v}$, $c_\text{s}$ are dimensionless coupling constants and ${\cal J}_f^\mu$, ${\cal J}_f$ are SM currents. Examples of such currents include
\begin{align}
{\cal J}_f^\mu &= \overline{f} \gamma^\mu f 
~,~
\overline{f} \gamma^\mu \gamma^5 f
\nla
{\cal J}_f &= \overline{f} f 
~,~
\overline{f} i \gamma^5 f
~,
\end{align}
where $f$ is a SM lepton ($\ell$) or nucleon ($N$). The DM annihilation rate near freeze-out is parametrically of size $\langle \sigma v \rangle \sim c_\text{v,s}^2 m_\x^2 / \Lambda^4$, implying that thermal freeze-out is consistent with the measured DM energy density for
\be
\Lambda \sim \left( c_\text{v,s} \, m_\x \right)^{1/2} \left( T_\text{eq} \, m_\text{pl} \right)^{1/4} \sim \order{10} \ \GeV \times \left( \frac{c_\text{v,s} \, m_\x}{100 \ \MeV} \right)^{1/2}
~,
\ee
where $T_\text{eq}$ is the temperature at matter-radiation equality and $m_\text{pl}$ is the Planck mass. 

If DM annihilates efficiently to SM matter, as it must for the thermal freeze-out mechanism to produce the observed abundance, then we can look for evidence of this annihilation in the universe at late times.  In the sub-GeV mass range, the tightest constraint comes from the annihilation of DM at eV-scale temperatures, shortly after recombination.  In the early universe, the annihilation products could have reionized atomic hydrogen, altering the CMB power spectrum. The total power injected has been constrained by Planck, leading to the constraint~\cite{Ade:2015xua,Aghanim:2018eyx}
\be
\langle \sigma v \rangle_{T\sim \eV} \lesssim \langle\sigma v \rangle_{T \sim m_\x} \left(\frac{m_\x}{50 \ \GeV}\right)
~,
\ee
where $\langle \sigma v \rangle_T$ denotes the DM annihilation cross section to visible final states at a temperature $T$.
It follows that, for MeV$-$GeV scale thermal DM, the annihilation rate to visible final states at recombination ($T \sim \eV$) must be \emph{considerably} suppressed  (by $2-5$ orders of magnitude) relative to the annihilation near freeze-out ($T\sim m_\x$).

This is a revealing constraint, but not a problematic one.  Many of the generic models for thermal DM involve a significant suppression of the annihilation rate at low temperatures. This is typically due to one of two effects: velocity-suppression or population-suppression. Annihilations of fermionic DM coupled to the SM through spin-0 currents or Majorana DM coupled to the SM through spin-1 currents, as in Eq.~(\ref{eq:DMcurrents2}), are $p$-wave with $\langle \sigma v \rangle \propto v^2$ and hence are velocity-suppressed at the time of recombination. This is also the case for scalar DM coupled to the SM through a spin-1 current,  
\be
{\cal J}_D^\mu = i (\x^\dagger \partial_\mu \x - \x \partial_\mu \x^\dagger)
~.
\ee
``Population-suppression" arises in scenarios involving $U(1)_D$-breaking mass terms ($\delta m$) that are much smaller than $U(1)_D$-preserving mass terms ($m_D$). Such models involve the vector coupling of nearly-degenerate Majorana mass-states, $\x_{1,2}$,
\be
\label{eq:DiracCurrent}
{\cal J}_D^\mu = i \bar{\x}_1 \gamma^\mu \x_2
~.
\ee
If $\x_2$ is only slightly heavier than $\x_1$, with general mass eigenvalues $m_{1,2} \simeq m_D \mp \delta m$, the leading annihilation process during freeze-out involves the excited state, $\x_2$, which is thermally depopulated at later times but well-before recombination. 

Details of the DM model, such as the spin-type of the interaction and the structure of the DM mass matrix, can dramatically suppress signals relying on \emph{non-relativistic} DM scattering, such as direct detection.  Indeed, in many scenarios of DM lighter than a few GeV, the strong suppression of DM \emph{annihilation} at low velocities, as required by the CMB, is accompanied by a suppression of low-velocity scattering.  One virtue of accelerator-based searches is that they involve momentum transfers comparable to or greater than the DM mass.  Their sensitivity is therefore independent of these details, allowing a robust and inclusive test of the sub-GeV thermal relic scenario.

\subsection{Force mediators and UV-complete portals}

The higher-dimensional operators discussed above are a valid description of a given process if the kinematic scale of that process is small compared to the mass-scale of the interaction, $\Lambda$. As summarized in the introduction, parametrically small couplings, as expected in dark sectors, can complicate the understanding of when the EFT is valid. Indeed, weakly-coupled light mediators will, when integrated out, lead to an EFT with a high characteristic cutoff scale $\La$, e.g., as given in Eq.~(\ref{cutoff}), even though the mediator mass scale is parametrically smaller. 

Each of the EFT interaction channels above can be extended to a wider kinematic range by ``integrating in'' the mediator field at tree-level, so that schematically
\be
\Lag \supset 
g_\text{SM} \, V \cdot {\cal J}_f + g_{\x} \, V \cdot {\cal J}_D
~,
\ee
where $V$ is the mediator field.
In practice, while low-energy descriptions like this are possible for each of the EFT operators discussed above, not all can be rendered independent of unknown short-distance physics (namely ultraviolet or UV-complete) in a simple manner. In many cases, the interactions are secretly still higher-dimensional when accounting either for anomalies or simply the chiral structure induced by the SM electroweak sector. As is now well known, there are only three UV-complete interactions of this type that are consistent with the SM electroweak symmetry-breaking structure. They involve couplings to new dark photon ($A'$), dark scalar ($S$), or dark fermion ($N$) degrees of freedom, and 
are characterized as follows: $\Lag_{\rm vector} = \frac{\ep}{2} \, F_{\mu\nu} F'_{\mu\nu} = - \ep A'_\mu J_{\rm em}^\mu$, $\Lag_{\rm Higgs} = (A \, S + \lambda \, S^2) \, H^\dagger H$, and $\Lag_{\rm neutrino} = Y_N \, LHN$, in terms of SM photon, Higgs and lepton doublet fields. On general EFT grounds, if these dark photon, scalar, or fermion degrees of freedom are present, we expect the three renormalizable portals to provide the leading sensitivity to them.

The simplest and most viable way in which light DM can interact with the SM is through the mixing of any force that DM is charged under (``dark forces'') with the photon. The basic setup that realizes this possibility consists of a massive dark sector gauge boson $A'$, with general Lagrangian containing, 
\be
\label{eq:lagrangian}
\Lag \supset  \frac{\epsilon}{2} \, F^\prime_{\mu\nu} F^{\mu \nu} +  \frac{1}{2} \, \mAp^2 \, A_\mu^\prime  A^{\prime \mu}  +  e \, A_\mu \,   {\cal J}_\text{em}^\mu + g_D \, \Ap_\mu \, {\cal J}^\mu_D
\, .
\ee
Here,  $\epsilon$ is the kinetic mixing parameter, $A^\prime$  is the massive ``dark photon" of a broken $U(1)_D$ symmetry, $A$ is the SM photon, 
 $F^\prime_{\mu\nu}$ and  $F_{\mu\nu}$ are  the dark-photon and electromagnetic field-strength tensors, ${\cal J}_\text{em}^\mu$ is the electromagnetic current, ${\cal J}_D^\mu$ is the dark current, and $\mAp$ is the dark photon's mass.

If $\mAp \ll \order{100} \ \GeV$, the dark photon predominantly mixes with the SM photon 
so that the visible sector acquires a millicharge under $U(1)_D$,
\be
\Lag \supset \eps \, e \, \Ap_\mu \, {\cal J}^\mu_\text{em}
~.
\ee
In this setup, the $U(1)_D$ symmetry is broken by the mass of the $\Ap$. Thus, the effective theory analysis of different scenarios should include other sources of soft $U(1)_D$ breaking, which we include below.

General discussions of hidden sector physics often include other possible portals to the dark sector: e.g., the UV-complete Higgs or neutrino portals noted above, dark vectors coupled primarily to lepton or baryon numbers, or pseudoscalars coupled through the axion portal.  Before proceeding, we pause briefly to explain why the focus on dark photons is appropriate here. The more general case of dark vector mediators is UV-sensitive, and thus theoretically distinct, but phenomenologically identical to dark photon physics \emph{except} that the dark gauge boson couplings to SM particles are not proportional to charge.  This changes the relative sensitivity of different DM production/scattering experiments in the expected way (leptophilic models favor DM-electron scattering and electron-production over DM-nuclear scattering and hadron-production, while the opposite is true for leptophobic models) and introduces other constraints on the mediator-SM coupling strength that are beyond the scope of this work.  The most minimal neutrino portal is not a viable portal for thermal DM production, though right-handed neutrinos are a viable DM candidate in their own right.  Furthermore, predictive scenarios involving dark (pseudo-)scalars are strongly constrained by meson decay constraints~\cite{Krnjaic:2015mbs}. Such considerations motivate the dark photon vector portal as the most viable, and we will focus on that case in the first instance, but will aim to illustrate the expectations for the other viable mediation channels as well.

\section{Introducing the $y - R$ Plane}
\label{sec:Rintro}

The DM reach of accelerator experiments is often presented in terms of the canonical parameter, $y \equiv \eps^2 \alpha_D (m_\x / \mAp)^4$, where $\al_D = g_D^2/(4\pi)$. This is simply related to the EFT cutoff $\Lambda$ ($y= (16\pi^2 \alpha_\text{em})^{-1} (m_\x/\Lambda)^4 \simeq 0.9 \, (m_\x/\Lambda)^4$), and well-motivated from considerations of DM production in the early universe. For $\mAp \gtrsim \text{few} \times m_\x$, the DM annihilation rate near freeze-out is proportional to $\langle \sigma v \rangle \propto \alpha_\text{em} \, y / m_\x^2$, implying that the cosmological $\x$ abundance is in agreement with the observed DM energy density for
\be
\label{eq:yfo}
y \sim \frac{m_\x^2}{\alpha_\text{em} \, T_\text{eq} \, m_\text{pl}} \sim \order{10^{-10}} \times \left( \frac{m_\x}{100 \ \MeV} \right)^2
~,
\ee
where $T_\text{eq}$ is the temperature at matter-radiation equality and $m_\text{pl}$ is the Planck mass. Although this cosmologically-motivated range for $y$ is independent of $\mAp / m_\x$ for $\mAp \gtrsim \text{few} \times m_\x$, DM observables at low-energy accelerators often scale with different powers of $m_\x$ and $\mAp$. Furthermore, the particular scaling of such observables depends on, e.g., the hierarchy between the typical collisional center of mass energy, $E_\text{cm}$, and $\mAp$. As a result, when presenting the sensitivities of terrestrial experiments in the $y - m_\x$ plane, it has become standard convention to fix the dark photon-to-DM mass ratio to one or several representative values, such as $\mAp / m_\x = 3, 5, \text{etc.}$ In this section, we further explore the range of DM parameter space spanned by $y$ and $ R \equiv \mAp / m_\x$, including regimes where DM production proceeds through off-shell mediators, i.e., $\mAp \lesssim 2 m_\x$ and $\mAp \gtrsim E_\text{cm}$.

\begin{figure}[t!]
\centerline{\includegraphics[width=11cm]{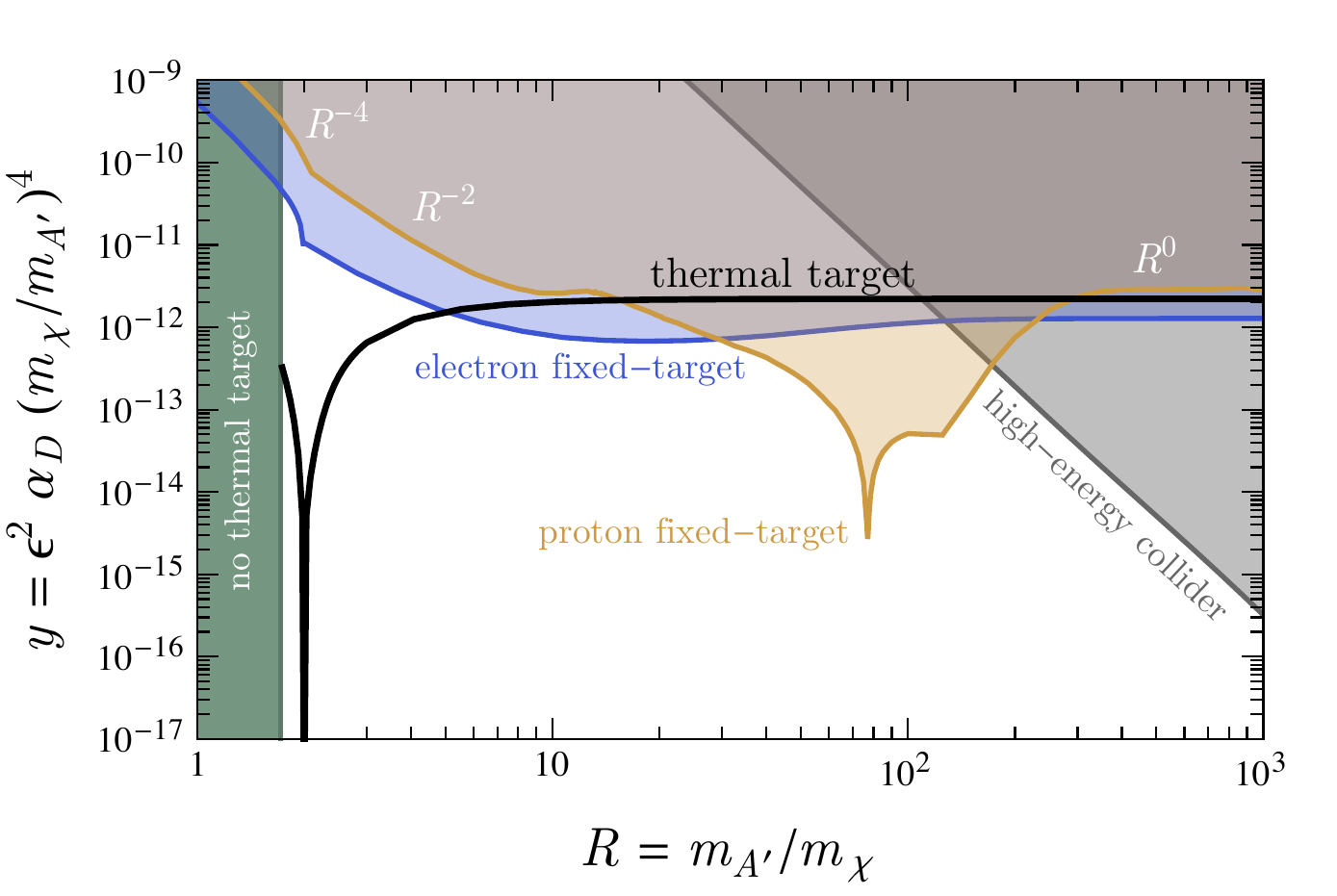}}
\caption{A schematic of the characteristic sensitivity scaling of electron fixed-target (blue), proton fixed-target (orange), and high-energy collider (gray) experiments in the  $y - R$ plane for a fixed DM mass $m_\x$. See the text for further details.} 
\label{fig:Rint}
\end{figure}

In Fig.~\ref{fig:Rint}, we schematically illustrate the characteristic sensitivity scaling of electron fixed-target (blue), proton fixed-target (orange), and high-energy collider (gray) experiments in the  $y - R$ plane for a fixed DM mass. 
Along the solid black line, the thermal freeze-out of $\x \bar{\x} \to A^{\prime *} \to f \bar{f}$ (where $f$ is an electrically charged SM fermion) in the early universe leads to an abundance of $\x$  that is in agreement with the observed DM energy density. For $R \gg 1$, the cosmologically-favored value of $y$ is given parametrically by Eq.~(\ref{eq:yfo}). Note that this region is independent of the particular value of the DM-dark photon coupling, $\alpha_D$. Near $R \simeq 2$, $\x \bar{\x} \to A^{\prime *} \to f \bar{f}$ is resonantly enhanced, allowing for an adequate annihilation rate near freeze-out with much smaller couplings~\cite{Griest:1990kh,Feng:2017drg}. For $R \lesssim 2$, the rate for $\eps$-independent reactions, such as $\x \x \to \Ap \Ap$ and $\x \x \Ap \to \Ap \Ap$, can significantly modify the process of thermal freeze-out, leading to an adequate abundance of $\x$ for significantly smaller values of $y$ (provided that $\alpha_D$ is sufficiently large and that the $\eps$-dependent decay, $\Ap \to e^+ e^-$, occurs sufficiently rapidly)~\cite{Pospelov:2007mp,DAgnolo:2015ujb,Cline:2017tka}. In this ``secluded'' regime, there is no sharp cosmological target and the cosmologically viable range of $y$ is extended to much smaller couplings. The sector of parameter space where the secluded annihilation $\x \x \to \Ap \Ap$ dominates over direct annihilations, $\x \x \to f \bar{f}$, is shown as the solid green region of Fig.~\ref{fig:Rint} (for a particular choice of $\alpha_D$).

In the remainder of this section, we will briefly discuss the qualitative behavior of the fixed-target and collider sensitivities in the $y-R$ plane, as shown in Fig.~\ref{fig:Rint}. More detailed discussion will be provided below in Secs.~\ref{sec:summary}-\ref{sec:sensitivity}.
\begin{itemize}
\item For $R \lesssim 2$, the on-shell decay $\Ap \to \x \x$ is kinematically forbidden, and hence DM production proceeds through reactions involving off-shell dark photons. Since in this case the typical accelerator center of mass energies are much greater than the dark photon mass, the DM production rate is independent of $\mAp$ and, hence, the corresponding sensitivities scale as $y_\text{reach} \propto R^{-4}$. 

\item For $R \gtrsim 2$ and $\mAp \lesssim E_\text{cm}$, DM production proceeds through on-shell production and decay of dark photons. In this region of parameter space, the signal yield for fixed $\epsilon$ at high-energy collider experiments is independent of $\mAp$, so that $y_\text{reach} \propto R^{-4}$, while the yield at fixed-target experiments typically scales as $\mAp^{-2}$, in which case the sensitivity to $y$ scales as $y_\text{reach} \propto R^{-2}$.   In electron fixed-target experiments, the finite size of the nucleus suppresses high-$\mAp$ production leading to a gradual flattening of the yield as $\mAp$ approaches $E_\text{cm}$.
For proton beams, there is additional structure in the $R$-dependence that emerges from thresholds in production from hadronic decays, and also resonant enhancements. This will be reviewed in Sec.~\ref{sec:production}.

\item For $\mAp \gtrsim E_\text{cm}$, collisions of electron or proton beams are unable to produce on-shell dark photons and DM production again occurs through an off-shell dark photon.  In this regime, the rate of DM production decouples as $\mAp^{-4}$. In this case, the dark photon can be integrated out from the low-energy theory and the sensitivity in the $y-R$ plane asymptotes to an $R$-independent value, $y_\text{reach} \propto \text{constant}$. \end{itemize}

\section{Summary of Experimental Signals at fixed-target Experiments}
\label{sec:summary}

In this section, we summarize the existing and proposed electron and proton fixed-target experiments whose capabilities for light DM detection will be analyzed in subsequent sections. Here, we provide a brief introduction and summary of detection signatures and provide further details of production modeling in the next section.

\begin{itemize}
\item {\it LDMX}
is a proposed experiment designed to search for signals of missing momentum in electron-nuclear fixed-target collisions of a high repetition rate ($\gtrsim 50 \text{ MHz}$) and energetic ($\gtrsim \text{GeV}$) electron beam~\cite{Akesson:2018vlm}. The momentum of every beam electron is measured by silicon trackers before and after it scatters in a thin $\sim 10 \%$ radiation-length tungsten target. This allows for a direct measurement of the momentum transfer that occurs in each electron-nuclear collision. Electromagnetic and hadronic calorimeters are placed downstream of this target region, in order to detect visible  activity, such as energy deposition from the recoiling beam electron and any other charged or neutral SM particles. DM that is produced in the electron-nuclear collision is characterized by a large energy loss and momentum exchange of the beam electron and the absence of measured activity in the downstream calorimeters, aside from the soft recoiling electron. The projected reach will be shown as blue lines/regions (see Figs.~\ref{fig:yR1}-\ref{fig:yRthermal} in Sec.~\ref{sec:sensitivity}).

\item {\it NA64}
is an existing experiment that searches for missing energy signals in electron-nuclear fixed-target collisions, utilizing the 100 GeV secondary electron beam at the CERN SPS on a thick lead target. Its setup is akin to that of the proposed LDMX configuration, without the downstream tracking sufficient to accurately measure missing momentum. Instead, DM that is produced in beam-target collisions registers as missing energy, which is inferred from the electromagnetic shower of the recoil electron in a downstream electromagnetic calorimeter. In this work, we recast the recently published limits from the NA64 collaboration,  corresponding to a total of $2.84 \times 10^{11}$ electrons on target (EOT) and background levels of $\lesssim 1$ event~\cite{NA64:2019imj}. The ensuing limits will be shown as solid gray (see Figs.~\ref{fig:yR1}-\ref{fig:yRthermal} in Sec.~\ref{sec:sensitivity}).

\item {\it BDX}
is a proposed DM beam dump experiment at JLab~\cite{Battaglieri:2016ggd}. It takes advantage of the high-current 11 GeV CEBAF electron beam on a thick aluminum target and is expected to acquire $10^{22}$ EOT within a year of parasitic running downstream of JLab Hall-A. DM that is produced in the electron-nuclear collisions is detected upon scattering in a shielded electromagnetic calorimeter placed $\sim 25 \ \m$ downstream of the target. This detector consists of $\sim 800$ CsI(Tl) crystals over an equivalent volume of a $\sim$ cubic meter and is sensitive to both DM-proton and DM-electron scattering. The projected reach will be shown as purple lines/regions (see Figs.~\ref{fig:yR1}-\ref{fig:yRthermal} in Sec.~\ref{sec:sensitivity}).

\item {\it SBND}
is under development as the near-detector component of the Fermilab short baseline neutrino (SBN) facility. Fed by the 8.9 GeV Booster Neutrino Beam (BNB), with $10^{21}$ protons on target (POT) expected for a 4 year run, SBND will be a 77 ton liquid argon (LAr) TPC, with a short baseline of 110 m. DM can be produced through proton-nucleon scattering in the beryllium target and absorber, via several channels including secondary $\pi^0$ and $\eta$ decays.  SBND is being designed to have low thresholds and will be sensitive to a variety of elastic and inelastic scattering signatures, both of neutrinos and potential light DM species. We focus on elastic electron scattering in this work, which allows for good neutrino background rejection via a tight forward angle cut, as utilized to good effect in the recent MiniBooNE off-target DM search run~\cite{MB1,MB2}. The projected reach will be shown as cyan lines/regions (see Figs.~\ref{fig:yR1}-\ref{fig:yRthermal} in Sec.~\ref{sec:sensitivity}). 

\item {\it NO$\nu$A}
is an operational long-baseline neutrino oscillation experiment, running from the 120 GeV NuMI beamline at Fermilab, and has now accumulated $\sim 10^{21}$ POT in both neutrino and antineutrino mode. Of most interest here is the 300 ton near detector, located 990 m from the carbon target. DM can be produced in the target and absorber as in the BNB, but the higher energy beam allows more significant production from, e.g., proton bremsstrahlung, and leads to a forward peaked DM distribution. We again focus on electron scattering, following the analysis in Ref.~\cite{deNiverville:2018dbu}, as the most effective means of limiting neutrino backgrounds. The projected reach is shown as yellow lines in Figs.~\ref{fig:yR1} and \ref{fig:yR2} (see Sec.~\ref{sec:sensitivity}).

\item {\it COHERENT}
is an expanding series of detectors at the SNS, focused on measuring coherent neutrino-nucleus scattering. The 1 GeV proton beam produces $10^{23}$ POT/year and a wide angle distribution of secondary mesons, and DM can be produced through $\pi^0$ decays or radiative pion capture. We utilize the recent analysis with the CENNS-10 liquid argon detector, a 27 kg shielded detector located at about 120 degrees to the beamline at a distance of approximately 28 m. The analysis placed a $1\sigma$ upper limit of 7.4 on the number of potential coherent elastic neutrino-nucleus scattering events~\cite{Akimov:2019rhz}, and we constrain light DM scenarios by assuming that all of these potential events were the product of DM scattering. This is shown as the gray region in the right panel of Fig.~\ref{LBplot} (see Sec.~\ref{sec:sensitivity}). The COHERENT collaboration recently published projections for the DM sensitivity of a proposed next generation ton-scale liquid argon detector~\cite{Akimov:2019xdj}.

\item {\it CCM}
(Coherent CAPTAIN-Mills) is a newly commissioned experiment using the 800 MeV proton beamline at LANSCE-Lujan, Los Alamos which can produce $10^{20}$ POT/yr. Similar to COHERENT at the SNS, it is designed to detect coherent neutrino scattering. However, the setup includes a 10 ton LAr detector placed 20 m from the tungsten target, which allows enhanced sensitivity, despite it being positioned at an angle of 120 degrees with respect to the beamline \cite{CCM}. We again focus on DM produced through $\pi^0$ decays and coherently scattering off of nuclei in the detector. The projected reach will be shown as brown lines (see Figs.~\ref{fig:yR1}-\ref{LBplot} in Sec.~\ref{sec:sensitivity}).

\end{itemize}

In addition to these fixed-target searches, we will also highlight in Figs.~\ref{fig:yR1}-\ref{fig:yRthermal} the existing constraints from a monophoton search at BaBar~\cite{Aubert:2008as,Lees:2017lec} and electroweak precision measurements at LEP (shaded gray)~\cite{Hook:2010tw,Curtin:2014cca}, as well as the projected sensitivity of a monophoton search at Belle II with 20 fb$^{-1}$ of integrated luminosity (red lines/regions)~\cite{Battaglieri:2017aum}. 

\section{Off-shell Production of Light Dark Matter at fixed-targets}
\label{sec:production}

\subsection{Off-shell production in electron fixed-targets}
We model the primary production of DM in electron-nuclear collisions in a modified version of {\tt MadGraph5}~\cite{mg}. This proceeds through on- or off-shell bremsstrahlung of an intermediate dark photon, i.e., $e N \to e N A^{\prime *} \to e N \x \x$. In this section, we briefly describe the computational procedure that we adopt throughout this work. 

Instead of generating independent signal event samples for each possible combination of the DM and dark photon mass, we generate one large statistical sample in the heavy-mediator/contact-operator limit for a discrete set of DM masses. As we will explain in more detail below, for each DM mass, the differential invariant mass spectrum can be rescaled accordingly for any dark photon mass. This allows for a more computationally efficient, but still accurate, calculation of the final signal yield.

As described above, the first step in our procedure is to generate large statistical samples of events for $e N \to e N \x \x$ in the contact-operator limit for a given experimental configuration, DM mass, spin, and interaction type. For instance, pseudo-Dirac DM particles that interact through a spin-1 coupling to electrons are modeled through a Lagrangian of the form
\be
\Lag_\text{EFT} \supset i g_\x \, \bar{\x}_1 \gamma_\mu \x_2 ~ \bar{e} \gamma^\mu e
~,
\ee
where $g_\x$ is a dimensionful coupling controlling the strength of the interaction, and similarly for other interaction types. We then bin, in the invariant mass of the DM pair $m_{\x \x}$, the subset of these events that pass the given experimental selection criteria. From this, we obtain the differential invariant mass spectrum, $d \sigma_{_\text{EFT}} /d \mxx$, in the heavy-dark-photon-limit. To obtain the invariant mass spectrum for any choice of dark photon mass or couplings, we rescale this distribution bin-by-bin by the squared propagator of an intermediate dark photon,
\be
\frac{d \sigma}{d \mxx} = \frac{(\sqrt{4 \pi \alpha_D} ~ \eps \, e / g_\x)^2}{(\mxx^2 - \mAp^2)^2 + (\mAp \Gamma_{\Ap})^2} ~ \frac{d \sigma_{_\text{EFT}}}{d \mxx}
~,
\ee
where above, $g_\x$ is the value of the contact-operator coupling that was chosen in making the initial distribution and $\Gamma_{\Ap}$ is the total dark photon width. In practice, we approximate the $\Ap$ to have an $\order{1}$ branching ratio to DM pairs, i.e., $\Gamma_{\Ap} \simeq \Gamma (\Ap \to \x \x)$, which is valid provided that $\alpha_D \gg \alpha_\text{em} \, \eps^2$. The total signal rate (accounting for experimental selection efficiencies) is calculated by integrating the rescaled distributions over all relevant values of the DM invariant mass,
\be
\sigma = \int d \mxx ~ \frac{d \sigma}{d \mxx}
~.
\ee

When simulating the DM production in this manner for each electron beam fixed-target experiment, we assume negligible background and utilize the following experimental parameters: 
\begin{itemize}
\item \textit{LDMX} - We adopt a beam energy of $E_\text{beam} = 16 \ \GeV$, a tungsten target thickness of 0.1 radiation-lengths, an integrated luminosity of $10^{16}$ EOT, and a $50\%$ signal efficiency. The signal region is defined by imposing that the energy of the recoil electron is $E_\text{e} \lesssim 0.3 \, E_\text{beam}$.
\item \textit{NA64} - In recasting the recently published limits from the NA64 collaboration, we adopt a beam energy of $E_\text{beam} = 100 \ \GeV$, a thick lead target, an integrated luminosity of $2.84 \times 10^{11}$ EOT, and a $50\%$ signal efficiency. The signal region is defined by imposing that the energy of the recoil electron is $E_\text{e} \lesssim 0.5 \, E_\text{beam}$. 
\item \textit{BDX} - We assume a beam energy of $E_\text{beam} = 11 \ \GeV$, a thick aluminum target, $10^{22}$ EOT, and a $20 \%$ signal detection efficiency. The target-detector distance is taken to be $25 \ \m$ and we model the front-face size and length of the CSI(Tl) detector as $50 \times 40 \ \cm^2$ and $250 \ \cm$, respectively. From our simulated sample of produced DM pairs, we analytically incorporate DM-electron scattering in the detector. The signal region is defined by imposing that the energy of the scattered electron is $E_\text{e} \gtrsim 500 \ \MeV$.
 \end{itemize}

\subsection{Off-shell Production in Proton fixed-targets}
The DM production rate at proton beam facilities is determined by $\si(pN\rightarrow A'^*+\cdots)$. The primary production modes have been analyzed in the literature (see Refs.~\cite{Batell:2009di,deNiverville:2011it,deNiverville:2012ij,Dharmapalan:2012xp,Batell:2014yra,Morrissey:2014yma,Kahn:2014sra,Gorbunov:2014wqa,Blumlein:2013cua}) and include:
\begin{itemize}
 \item {\it $\pi^0/\eta$ decay in flight} - The channel $\pi^0,\et \longrightarrow \gamma + A'^{*} (\rightarrow \x \x)$ is often the dominant mode. Given the 3-body decay distribution in the contact operator limit, we can write the full differential distribution (with $A'$ on- or off-shell)
 as 
\be
\frac{d\Gamma_{\pi^0 , \eta \rightarrow \gamma \x \x}}{dq^2} = \frac{m_{A^{\prime}}^4}{(q^2-m_{A^{\prime}}^2)^2+\mAp^2\Gamma_{A^{\prime}}^2} \left.\frac{d\Gamma_{\pi^0 , \eta \rightarrow \gamma \x \x}}{dq^2}\right|_{\rm EFT}
\, ,
\ee
where $q^2= m_{\x \x}^2$ 
  is the invariant mass of a DM pair, i.e., the dark photon virtuality.
 \item {\it Bremsstrahlung with resonant vector meson mixing} - For this channel $p +N \to p + N + A'^* (\rightarrow \x \x)$, standard approaches involving variants of the Weizsacker Williams approximation require that kinematic criteria such as $E_p, E_{A^{\prime}} \gg m_V$ be satisfied~\cite{dNCPR}, which are restrictive if $A'$ is highly off-shell. However, in practice this mode is significant only over the $m_{A^{\prime}}$ range for which the timelike form-factor of the radiated $A'$ is dominated by the vector resonances, e.g., $m_{A^{\prime}} \sim m_{\rho,\om}$. Therefore, for the DM masses of interest here, the on-shell approximation is sufficient.
 \item {\it Drell-Yan production from quark/gluon constituents} - Once the dark photon virtuality $q^2$ is above the characteristic hadronic scale, parton-level Drell-Yan processes $ q \bar{q}\rightarrow A'^* (\rightarrow \x \x)$ provide the relevant description. We have
 \be
 \frac{d\si}{dq^2}(pN\rightarrow A'^* (\rightarrow \x \x)+\cdots)= \sum_q  \mathcal{L}_{q\bar{q}}(q^2) \, \frac{m_{A^{\prime}}^4}{(q^2-m_V^2)^2+ m_{A^{\prime}}^2\Ga_{A^{\prime}}^2} \, \frac{d\si}{dq^2}(q\bar{q}\rightarrow \x \x)|_{\rm EFT}+\cdots
 \, ,
 \ee
 where $\mathcal{L}_{q\bar{q}}(q^2)$ is the parton luminosity.
\end{itemize}

We focus our attention on meson decays and bremsstrahlung as the two channels expected to dominate the low-mass regime to which proton beam dump experiments are most sensitive. We make use of the following parameters when simulating the signal at each experiment:
\begin{itemize}
 \item \textit{SBND} - The Sanford Wang parameterization~\cite{AguilarArevalo:2008yp} was sampled to generate mesons produced using the BNB, with $\pi^0$s simulated by the the mean of a $\pi^+$ and $\pi^-$ distribution and $\eta$'s from the $K^0$ distribution. The overall production rates were estimated to be $N_{\pi^0} = 0.9 \times$POT and $N_{\eta} = N_{\pi^0}/30$. For bremsstrahlung, we placed limits on $z = E_{A^{\prime}}/E_p$ of $z\in[0.3,0.7]$, adjusting the limits on $z$ such that $z E_p, (1-z) E_p > 3 m_{A^{\prime}}$ and the transverse ${A^{\prime}}$ momentum satisfies $p_T < 0.1\,\mathrm{GeV}$.

 For the NCE electron scattering, we assume that all neutrino backgrounds can be rejected with a $\cos \theta>0.99$ cut on the scattering angle of the recoil electron, and a sensitivity limit can be placed with 2.3 events. For the inelastic $\pi^0$ scattering case, we make no cuts on the recoil particles but assume that at least 60 events are required to overcome existing backgrounds. The event rate is calculated for $10^{21}$ POT with a detection efficiency of 60\% in both cases.
 
 \item \textit{\Nova} - Both the $\pi^0$ and $\eta$ production distributions from the NuMI beamline were simulated by sampling the BMPT distribution~\cite{Bonesini:2001iz}. The $\pi^0$ production rate $N_{\pi^0}$ was conservatively estimated at 1 per POT with $N_\eta = 0.078 N_{\pi^0}$. For bremsstrahlung, we put limits of $z \in [0.1,0.9]$ and $p_T < 1\,\mathrm{GeV}$, once again limiting $z$ such that $z E_p, (1-z) E_p > 3 \mAp$.
 
  Only NCE-like electron scattering was considered for NO$\nu$A. We follow the treatment of Ref.~\cite{deNiverville:2018dbu}, selecting recoil electrons with $E\in[0.5,5]\,\mathrm{GeV}$ and $E\theta^2 < 0.005\,\mathrm{GeV}\,\mathrm{rad}^2$. We place a limit on 41 scattering events with a detection efficiency of $50\%$.
 
 \item \textit{COHERENT} and \textit{CCM} - The Burman-Smith parametrization~\cite{Burman:1989ds} was sampled to generate mesons produced along the SNS or LANSCE beams, with $\pi^0$s simulated by a $\pi^+$ distribution due to the lack of a $\pi^-$ distribution at these energies. For COHERENT, we take $N_{\pi^0}=0.06\times\mathrm{POT}$, while for CCM we take it to be $N_{\pi^0}=0.0425\times\mathrm{POT}$.
 
 For COHERENT, we considered coherent DM scattering off of Liquid Argon with a minimum energy cut of 80 keVnr,\footnote{This cut is expected to improve to 20 keVnr in the production run.} and placed a $1\sigma$ limit on 7.4 coherent DM scattering events over a $4.2\times10^{22}$ POT run. At CCM, we adopted a minimum energy cut of 10 keVnr and placed a limit on 10 coherent DM scattering events. In both cases, we assume $50\%$ efficiency.
 
\end{itemize}

The projected sensitivity contours in Fig.~\ref{fig:yRthermal} were generated by simulating the expected signal at each experiment for 2500 combinations of $m_{A^\prime}$ and $m_\chi$.

\section{DM Sensitivity contours at fixed-targets}
\label{sec:sensitivity}

We now combine the production modeling above with the relevant detection signatures at each experiment to produce sensitivity contours. The contours are chosen to reflect either known or estimated backgrounds, as outlined in each case. We will present two classes of figures, with the goal of exhibiting the full parameter space of light DM coupled to the SM through the vector portal. 

\subsection{The $y-R$ plane}
\label{sec:yR}

\begin{figure}[t]
\centerline{\includegraphics[width=10cm]{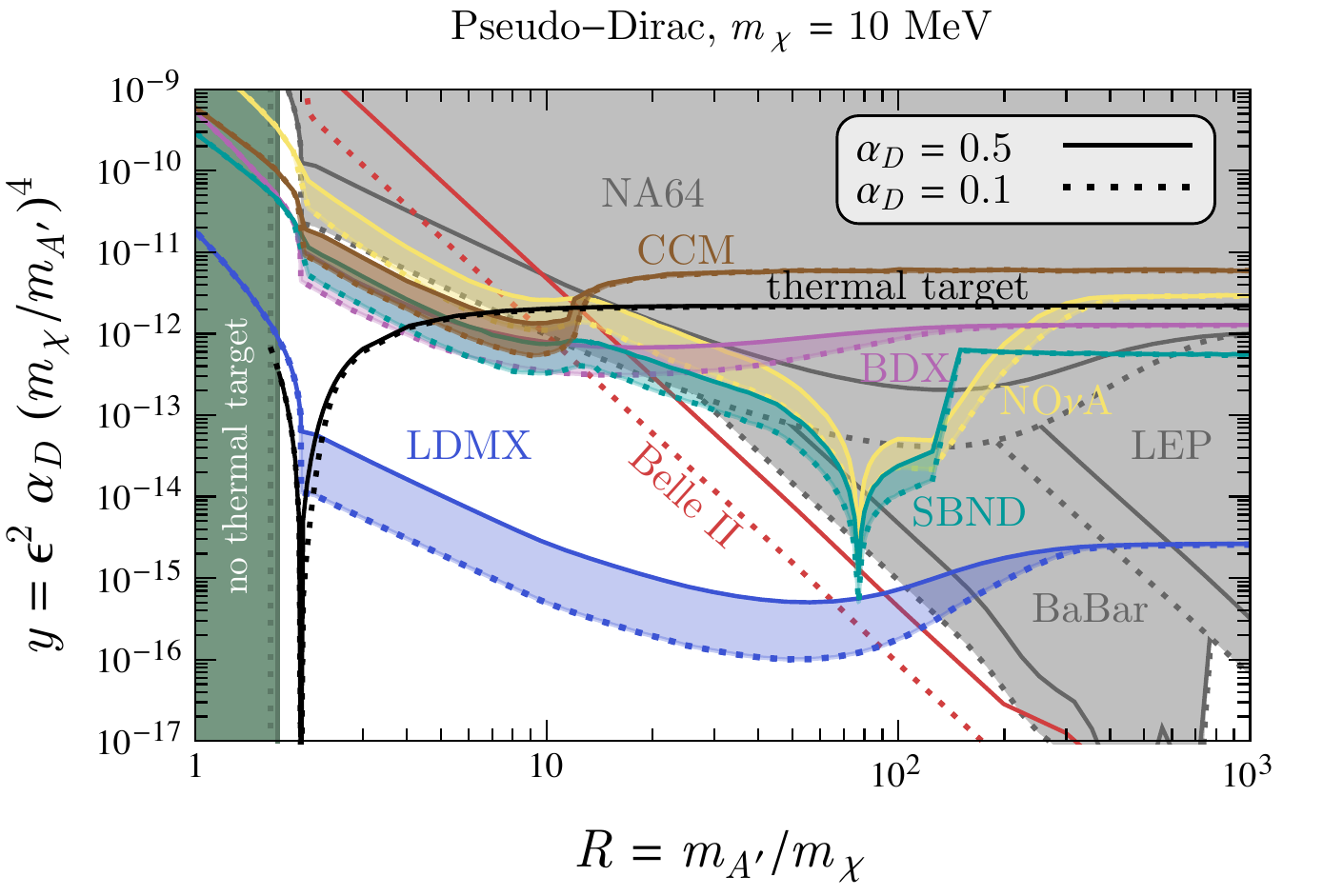}\includegraphics[width=10cm]{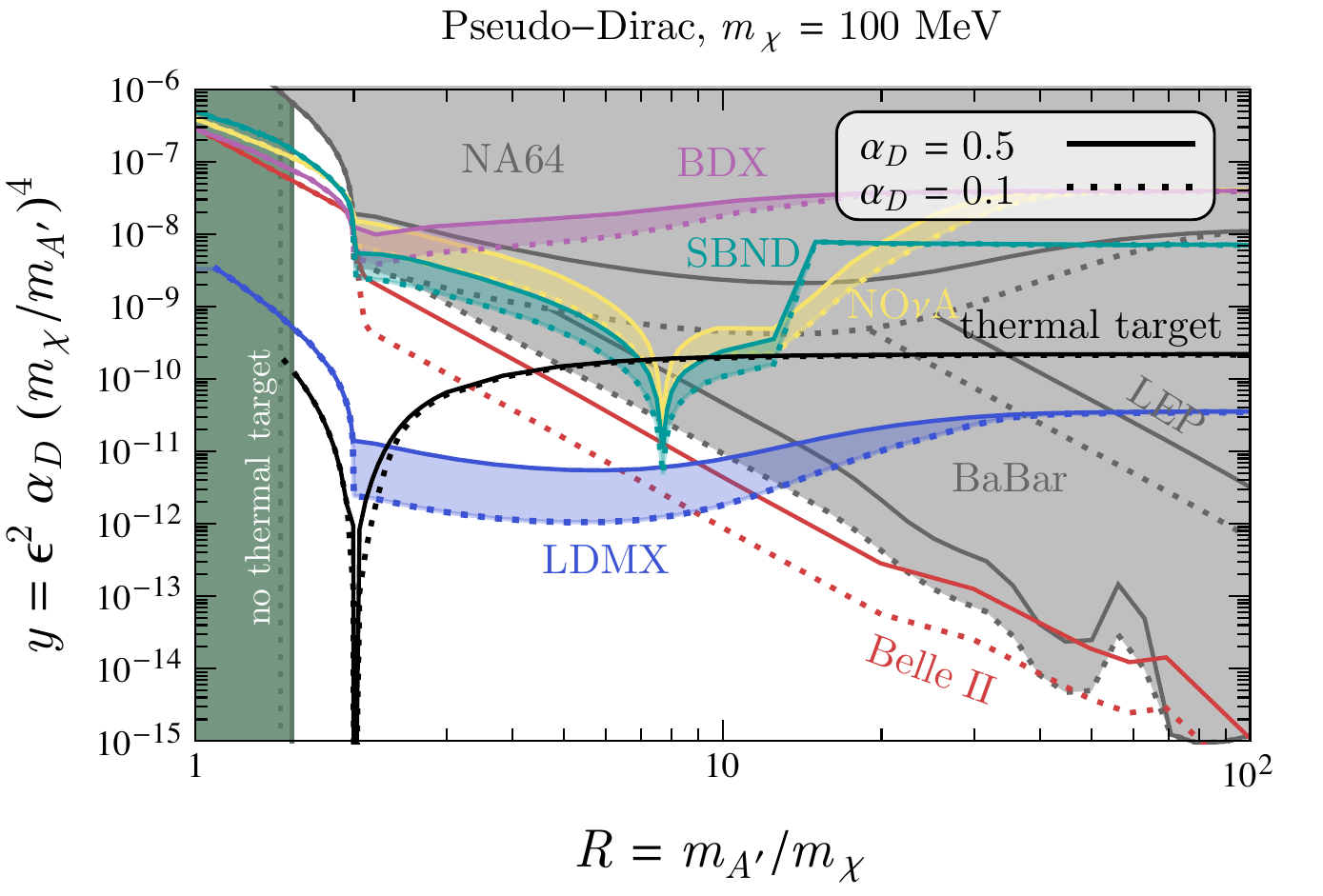}}
\caption{Existing constraints (shaded gray) and projected sensitivities (color) of various accelerator experiments to light dark matter in the $y-R$ plane are shown for fixed dark matter masses of $10 \ \MeV$ (left panel) and $100 \ \MeV$ (right panel). See Sec.~\ref{sec:summary} for a comprehensive summary. In each case, we consider two representative values of the hidden sector gauge coupling, $\alpha_D = 0.1, 0.5$, as shown by dotted and solid lines, respectively. For concreteness, we focus on a model of dark matter consisting of a nearly-degenerate pseudo-Dirac pair; the results for Majorana or scalar dark matter are qualitatively very similar (see the right panel of Fig.~\ref{fig:yR2}). Along the black lines, the abundance of $\x$ agrees with the observed dark matter energy density. For $R \lesssim 2$, the shaded green region corresponds to dark photon-to-dark matter mass ratios for which secluded annihilations dominate over direct annihilations to Standard Model particles. In this case, there is no sharp cosmological target in parameter space.} 
\label{fig:yR1}
\end{figure}

In this section, we present the fixed target and collider sensitivities in the $y-R$ plane, where $R\equiv \mAp / m_\x$. For a fixed choice of DM mass $m_\ch$, this allows us to explore the extrapolation in sensitivity from the fully off-shell EFT regime at large masses ($\mAp \gg E_{\rm cm}$) to the opposite regime where on-shell decays may enhance the rate, or resonant effects become important. In Fig.~\ref{fig:yR1}, we show results for pseudo-Dirac DM and two choices of $m_{\ch}$ ($10 \ \MeV$ and $100 \ \MeV$) and DM-dark photon coupling ($\al_D = 0.1, 0.5$). Many of the qualitative features were already discussed briefly in Sec.~\ref{sec:Rintro}. We summarize some of the physical features below:

\begin{itemize}
\item {\it Thermal target} - The parameters required to ensure the full DM relic abundance from freeze-out is shown as a solid/dotted black line. The annihilation process, $\x \bar{\x} \to A^{\prime *} \to f \bar{f}$, is resonantly enhanced near $R=2$. For $R \lesssim 2$, ``secluded" processes such as $\x \bar{\x} \to \Ap \Ap$ lead to an adequate abundance for much smaller couplings to the SM and there is effectively no sharp thermal target (shaded green).  
\item {\it EFT region} - The $R \gg 1$ regions of the figures show that the sensitivity of fixed-target experiments asymptotes to a fixed value. This occurs beyond the value of $R$ at which the production becomes predominantly off-shell and well-approximated by a contact operator. Note that the contours for LEP (and, in Fig.~\ref{fig:yR1} (right), BaBar and Belle II) do not flatten within the range of the plot, as the $A'$ can still be produced on-shell for the parameter range shown. 
\item {\it Resonant production} - For proton fixed-target experiments, bremsstrahlung of the dark photon mediator (and thus DM production) is resonantly enhanced when $\mAp$ is close to the mass of one of the SM vector resonances, such as the  $\rho$ meson.
\item {\it Light off-shell window} - For 
$R < 2$, DM production necessarily involves off-shell dark photons, leading to the reduced sensitivity that is apparent in both panels of Fig.~\ref{fig:yR1}.
\item {\it R=3} - In much of the recent literature, the reach of various experiments is compared after fixing $R=3$. We observe that this is relatively conservative, but does not fully illustrate the resonant freeze-out region near $R=2$ or the off-shell freeze-out region of Eq.~(\ref{eq:yfo}).
\end{itemize}

\begin{figure}[t]
\centerline{
\includegraphics[width=10cm]{Figures/Dirac_mX_10MeV.pdf}
\includegraphics[width=10cm]{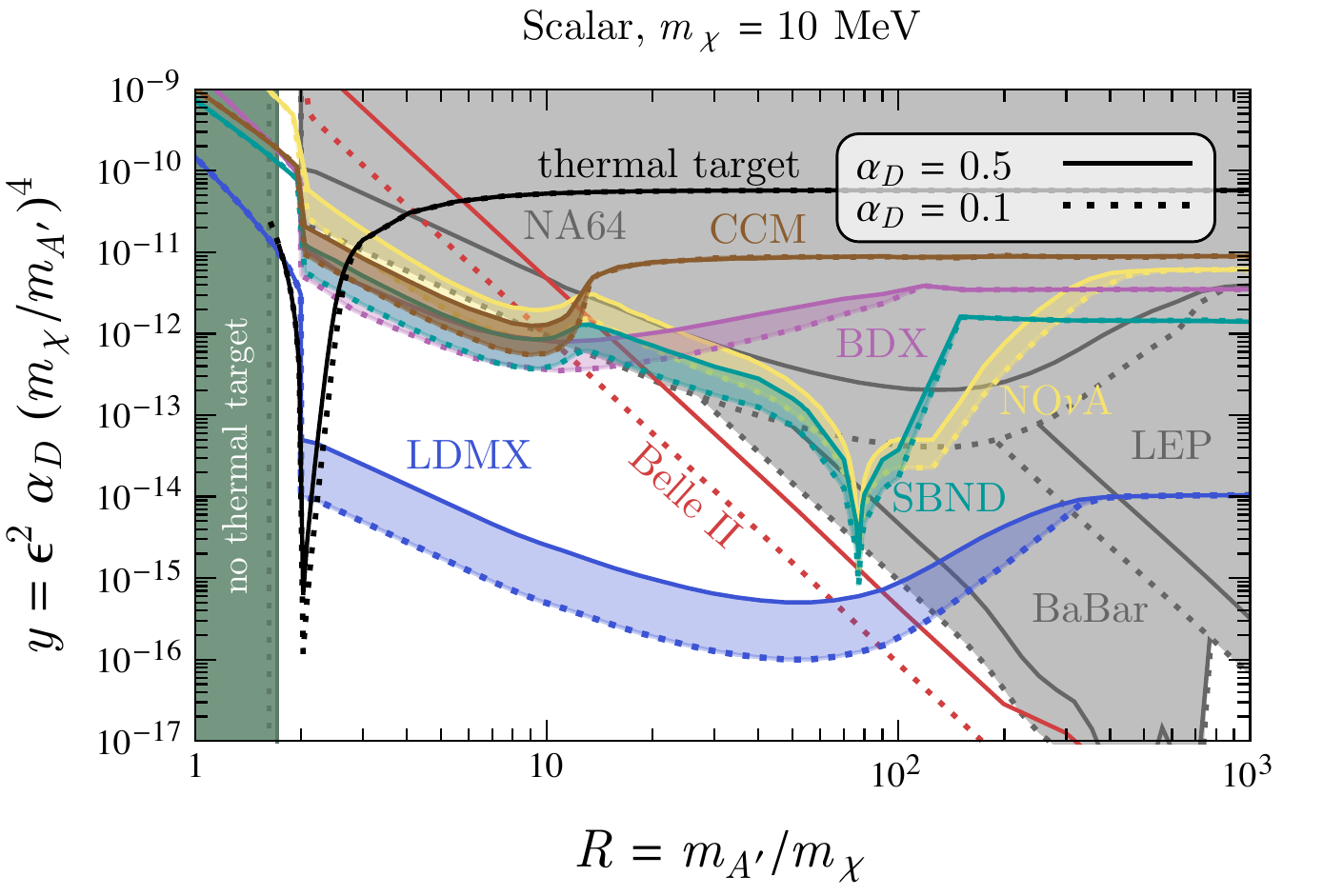}
}
\caption{As in Fig.~\ref{fig:yR1}, we show existing constraints (shaded gray) and projected sensitivities (color) of various accelerator experiments to light dark matter in the $y-R$ plane, but for a fixed dark matter mass of $10 \ \MeV$ and for models of pseudo-Dirac (left panel) \emph{and} scalar dark matter (right panel). The left panel is repeated from Fig.~\ref{fig:yR1} to aid visual comparison between the two cases.}
\label{fig:yR2}
\end{figure}

In the right-panel of Fig.~\ref{fig:yR2}, we show the results of a similar set of calculations for scalar DM, fixing $m_\x = 10 \ \MeV$. To aid the visual comparison between the two models, we show again the results for pseudo-Dirac DM in the left-panel. We note that the accelerator reach for these two models is very comparable, aside from minor differences when the dark photon is highly off-shell, due to the distinct spin-structure of the DM-dark photon couplings. While the qualitative features of the thermal target are similar for pseudo-Dirac and scalar DM, DM annihilations to SM fermions is $p$-wave suppressed during freeze-out for scalar DM. As a result, the thermal target for scalar DM is shifted to slightly larger SM couplings.

\begin{figure}[t]
\centerline{\includegraphics[width=10cm]{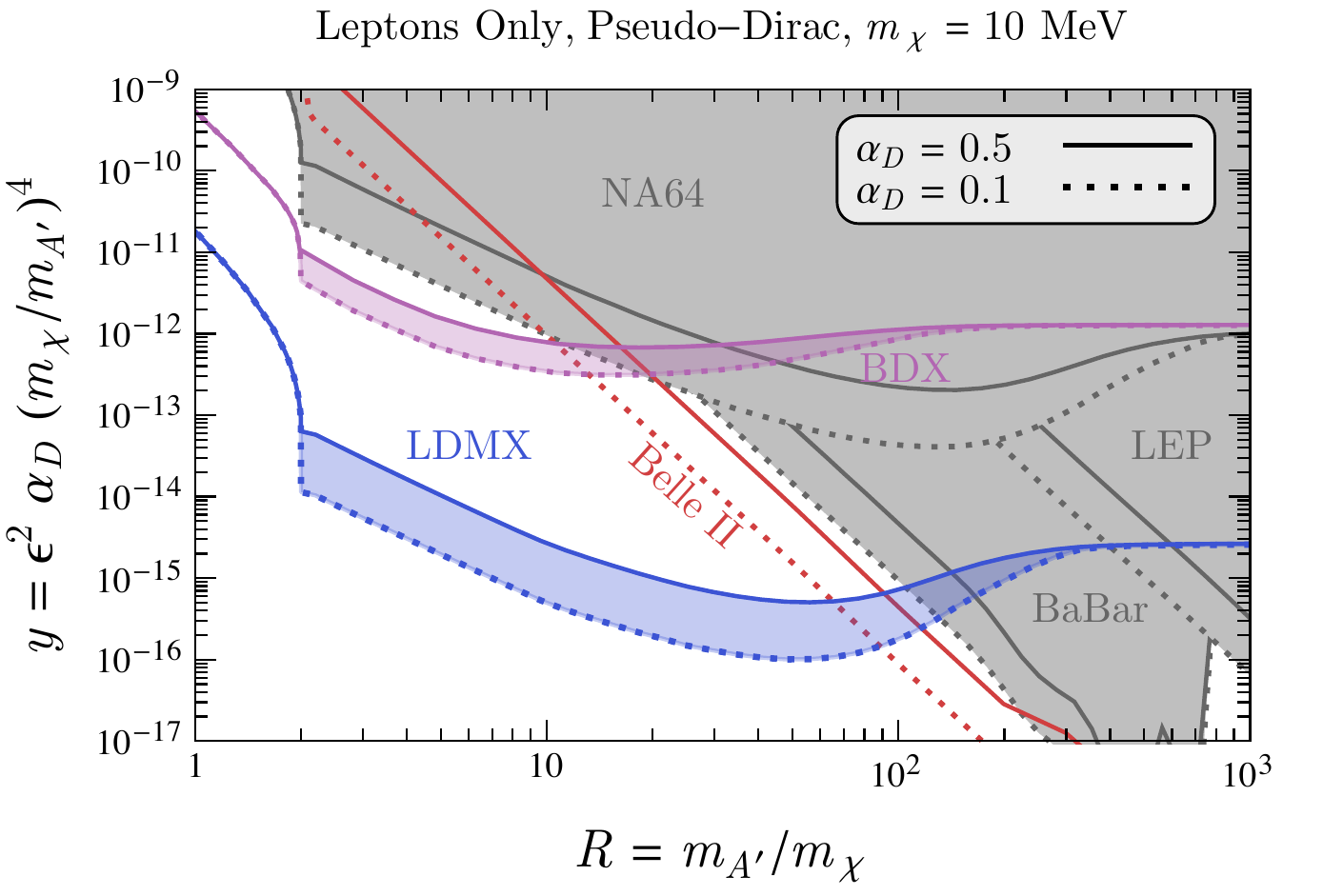}\includegraphics[width=10cm]{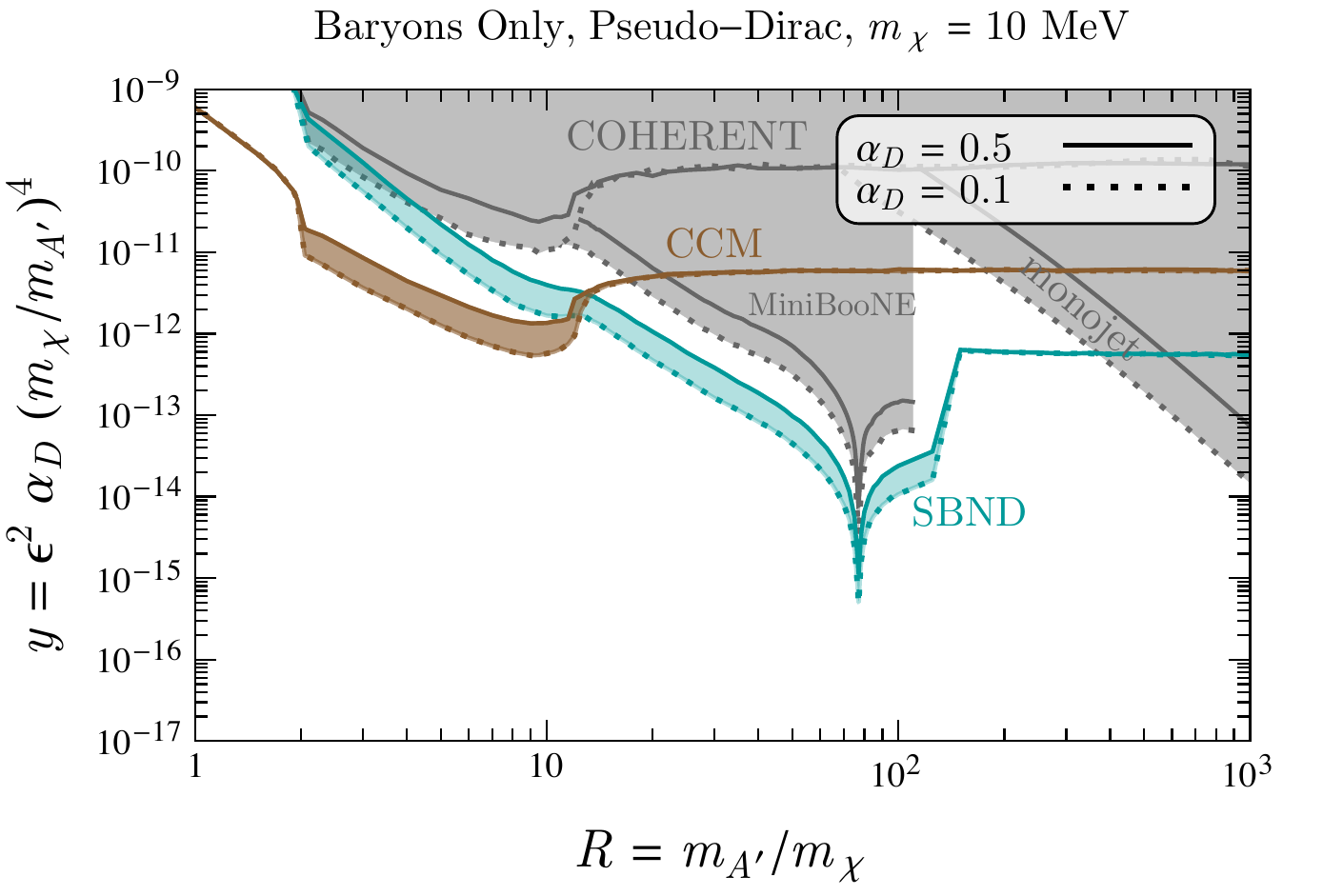}}
\caption{As in Fig.~\ref{fig:yR1}, existing constraints (shaded gray) and projected sensitivities (color) of various accelerator experiments to light dark matter in the $y-R$ plane are presented, but for a fixed dark matter mass of $10 \ \MeV$ and only showing signals that depend exclusively on leptonic (left panel) or hadronic (right panel) couplings.
Note that the same model is constrained in each case and that these figures simply indicate the sensitivity of these experiments. Fully consistent models with predominant leptonic or hadronic couplings require further model-building and may be subject to a number of other stringent constraints (see the text for further details). For example, leptophobic models coupled through the anomalous baryon current are subject to strong constraints from meson decays~\cite{Dror:2018wfl,Dror:2017ehi,Dror:2017nsg}.} 
\label{LBplot}
\end{figure}

Separately, in Fig.~\ref{LBplot}, we show the reach of experiments that depend only on leptonic or hadronic couplings, respectively. Compared to Figs.~\ref{fig:yR1} and \ref{fig:yR2}, in the right panel of Fig.~\ref{LBplot} we also include existing constraints from a DM search at MiniBooNE~\cite{Aguilar-Arevalo:2018wea} and monojet searches at high-energy proton colliders, such as the Tevatron and the LHC~\cite{Shoemaker:2011vi}. We choose to present the complementary sensitivity in this way, rather than working with fully UV-complete models of this type, e.g., those involving mediators such as $L_\mu - L_\tau$ or $B - 3L_\ta$~\cite{Berlin:2018bsc}, partly for simplicity, but also because such models entail a number of other more model-dependent constraints. It is important to note that in any specific model, there may be additional constraints on this parameter space that will be relevant. A precise specification of the model is also necessary to determine whether a thermal target lies in unconstrained parameter space.


The sensitivity contours of the electron beam experiments can be understood by analyzing the production channels for each experiment that are relevant as a function of $R$.
The existing constraint from an electron-beam missing-energy search at NA64 is shown as shaded gray in Figs.~\ref{fig:yR1} and \ref{fig:yR2}. The projected sensitivity of a missing momentum search at LDMX is also shown as shaded blue. For $\mAp \ll m_\x$, the on-shell decay $\Ap \to \x \x$ is kinematically forbidden, and hence DM production proceeds through an off-shell dark photon, i.e., $e N \to e N A^{\prime *} \to e N \x \x$. Therefore, the expected number of signal events scales as $N_\text{signal} \propto \alpha_D \eps^2$. Instead, for dark photons above the DM mass-threshold, $\mAp \gtrsim 2 m_\x$, but lighter than the typical center of mass energy of an electron-nucleus collision, $\mAp \lesssim E_\text{cm}$, DM production proceeds through standard on-shell processes. In this case, if the dark photon has an $\order{1}$ branching ratio to DM pairs, the signal rate scales as the on-shell $\Ap$ production rate, i.e., $N_\text{signal} \propto \eps^2 / \mAp^2$~\cite{Bjorken:2009mm}. For dark photons much heavier than the typical center of mass energy, $\mAp \gg E_\text{cm}$, DM production once again proceeds through off-shell processes, but the rate is now further suppressed by the dark photon mass, $N_\text{signal} \propto \alpha_D \eps^2 / \mAp^4$. Therefore, for each of these various mass regimes, for a fixed DM mass the sensitivity of NA64 and LDMX in the $y-R$ plane scales as
\be
\label{eq:yNA64}
y_\text{reach} (\text{NA64/LDMX})  \propto
\begin{cases}
R^{-4}, & (\mAp \lesssim 2 m_\x)
\\
\alpha_D \, R^{-2}, & (\mAp \gtrsim 2 m_\x)
\\
\text{constant}, & (\mAp \gg E_\text{cm})
~.
\end{cases}
\ee
For NA64 and LDMX, the typical center of mass energy in the electron-nucleus collisions is roughly $5 \ \GeV$ and $2 \ \GeV$, respectively. The $R$ and $\alpha_D$ scaling of the different mass regions of Eq.~(\ref{eq:yNA64}) can be seen directly in the behavior of the NA64 and LDMX contours of Figs.~\ref{fig:yR1} and \ref{fig:yR2}. In particular, taking $E_\text{cm} \sim 5 \ \GeV$ and $m_\x = 10 \ \MeV$ implies that NA64's sensitivity in $y$ should asymptote to a constant value for $\mAp / m_\x \gtrsim 500$. 

The projected sensitivity of the BDX experiment is shown as shaded purple in Figs.~\ref{fig:yR1} and \ref{fig:yR2}. The $\alpha_D$ and $R$ scaling in the $y-R$ plane can be understood in a manner very similar to the discussion regarding NA64 and LDMX above. However, the signal yield needs to be supplemented with the scattering rate for $\x e \to \x e$ in the downstream detector. The total cross section for the scattering process scales as $\sigma (\x e \to \x e) \propto \alpha_D \, \eps^2 (2 m_e E_\text{th} + \mAp^2)^{-1} (2 m_e E_\text{beam} + \mAp^2)^{-1}$. $E_\text{th} \sim 500 \ \MeV$ is the minimum threshold energy of the detected scattered electron, so that $m_e E_\text{th} \sim (10 \ \MeV)^2$. Similar to Eq.~(\ref{eq:yNA64}), we therefore have
\be
\label{eq:yBDX}
y_\text{reach} (\text{BDX})  \propto
\begin{cases}
R^{-4}, & (\mAp \lesssim 2 m_\x, \sqrt{m_e E_\text{th}})
\\
R^{-3}, & (\sqrt{m_e E_\text{th}} \lesssim \mAp \lesssim 2 m_\x)
\\
\alpha_D^{1/2} \, R^{-2}, & (\mAp \gtrsim 2 m_\x, \sqrt{m_e E_\text{th}})
\\
\text{constant}, & (\mAp \gg E_\text{cm})
~.
\end{cases}
\ee

For the proton beam fixed-target experiments, CCM, SBND, and \Nova, the sensitivity can be understood in a similar manner to BDX, as the production rate needs to be convoluted with the scattering cross section on either electrons (SBND, \Nova) or nuclei (CCM); thus $N_{\rm signal} \propto y^2$. The dominant production modes for $2m_\ch < m_{\pi/\eta}$ are via pseudoscalar meson decay, and so $N_{\rm signal} \sim \ep^2 \al_D$ for $\mAp < 2 m_\ch$, $N_{\rm signal} \sim \ep^2$ for $\mAp > 2 m_\chi$ provided that the dark photon is produced on-shell, or $N_{\rm signal} \sim \al_D \ep^2/\mAp^4$ if $\mAp > m_{\pi/\eta}$ and the dark photon is off-shell. The scaling of the scattering cross section is as above for BDX (or replacing $m_e$ with $m_N$ for nucleon scattering). For SBND and \Nova, the best reach at low mass comes from electron scattering, although the contours on the right of Fig.~\ref{LBplot} rely on nucleon scattering. We find that the reach in $y$ is as follows:
\be
\label{eq:yproton}
y_\text{reach} (\text{SBND/\Nova/CCM})  \propto
\begin{cases}
R^{-4}, & (\mAp \lesssim 2 m_\x, \sqrt{m_{e/N} E_\text{th}}, m_{\pi/\eta})
\\
R^{-3}, & (\sqrt{m_{e/N} E_\text{th}} \lesssim \mAp \lesssim 2 m_\x \lesssim m_{\pi/\eta})
\\
\alpha_D^{1/2} \, R^{-2}, & (m_{\pi/\eta} \gtrsim \mAp \gtrsim 2 m_\x, \sqrt{m_{e/N} E_\text{beam}})
\\
\text{constant}, & (\mAp \gg m_{\pi/\eta})
~.
\end{cases}
\ee
The primary exception to this scaling is near the $\mAp \sim m_{\rho/\om}$ resonance region, where production via proton bremsstrahlung is resonantly enhanced through mixing with vector mesons. This resonant peak is apparent in Figs.~\ref{fig:yR1} and \ref{fig:yR2} for SBND and \Nova, but the beam energy is too low at CCM to access the resonant regime. The $1/R^2$ scaling for CCM, SBND and NO$\nu$A is apparent for $\mAp < m_\pi$, and similarly (with reduced reach) for SBND and \Nova\ for $m_\pi < \mAp < m_\eta$. Just above this scale, the reach is enhanced by resonant mixing with the $\rho$ and $\om$ mesons.

\begin{figure}[t]
\includegraphics[width=8cm]{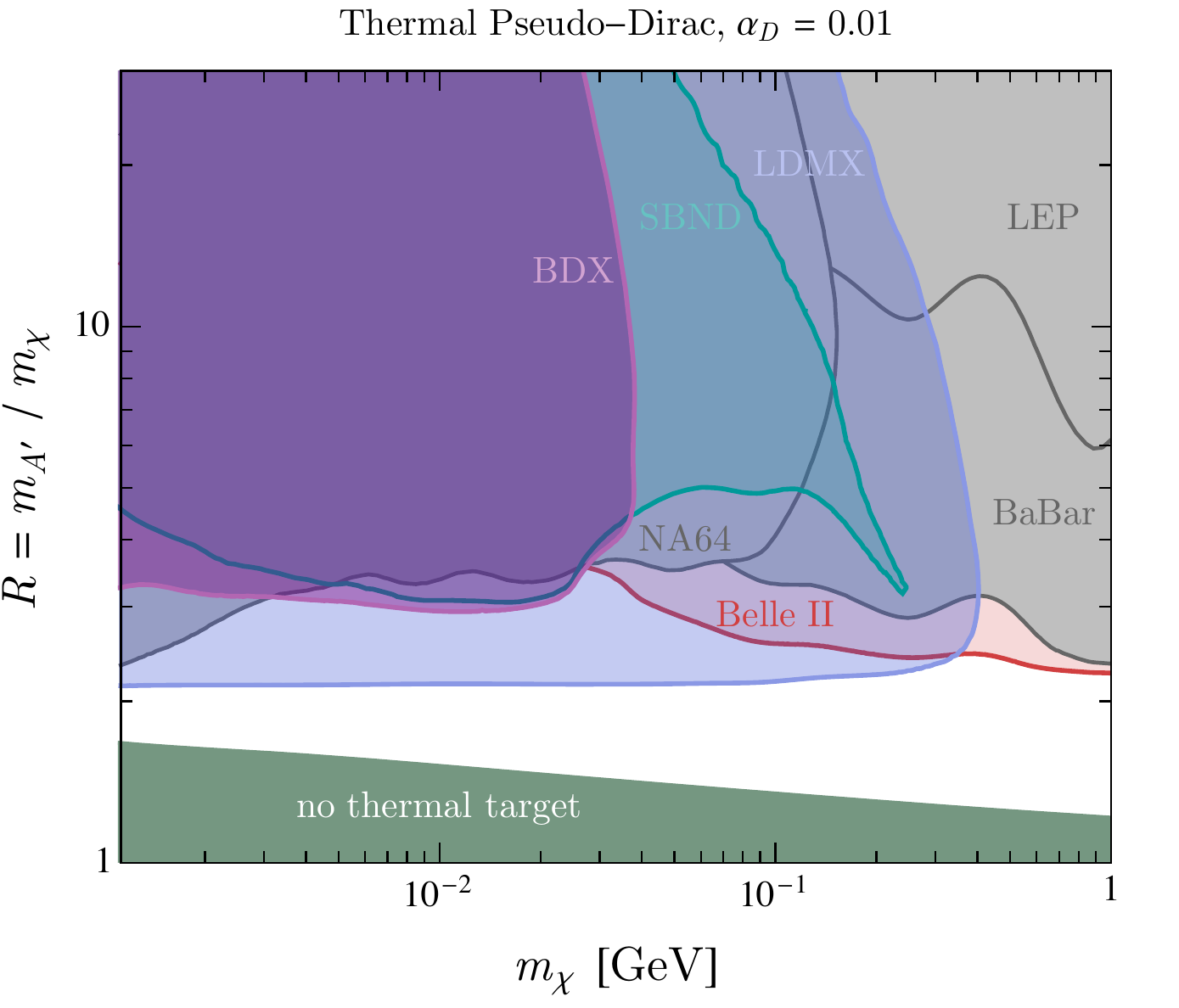}
\includegraphics[width=8cm]{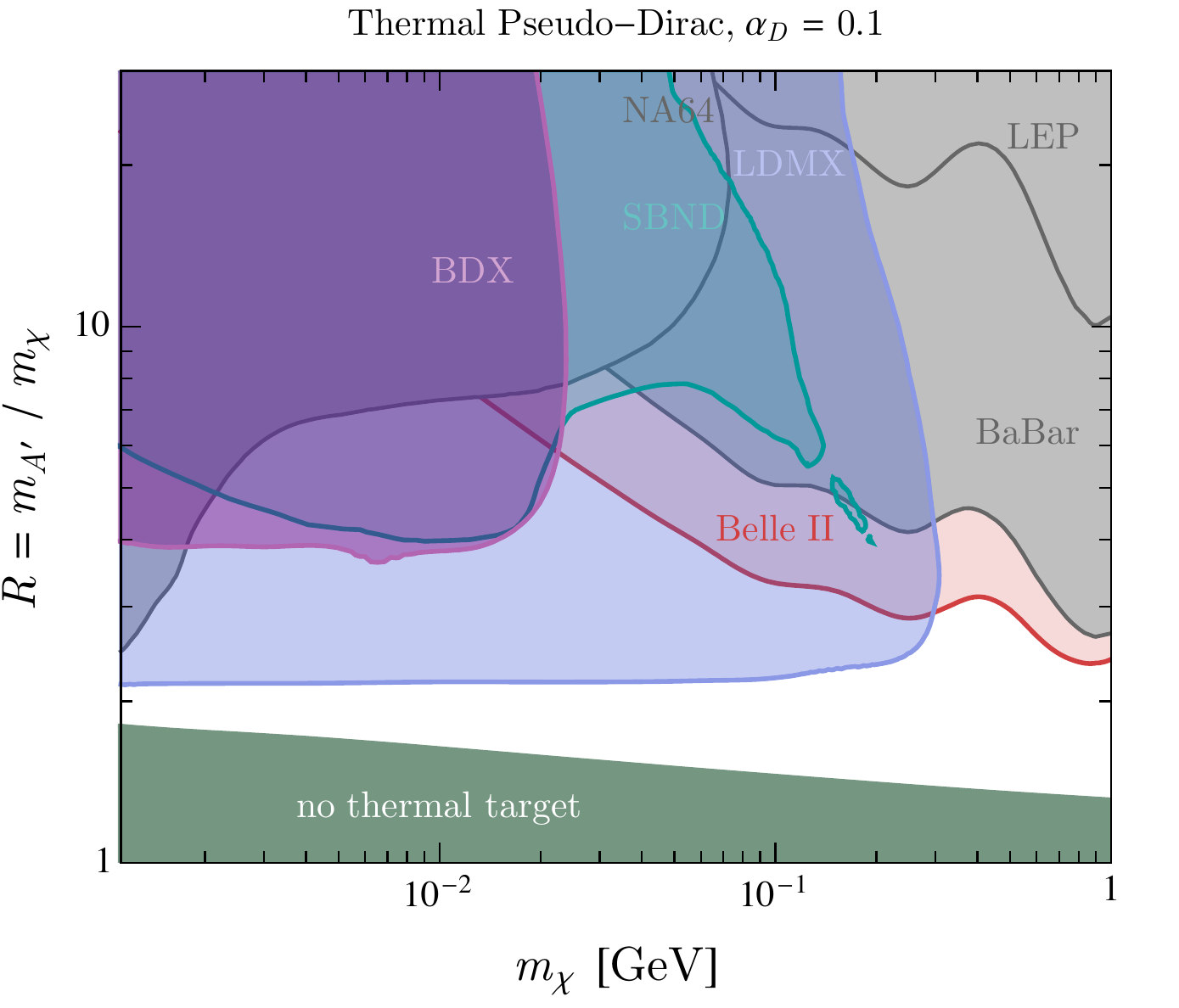}
\includegraphics[width=8cm]{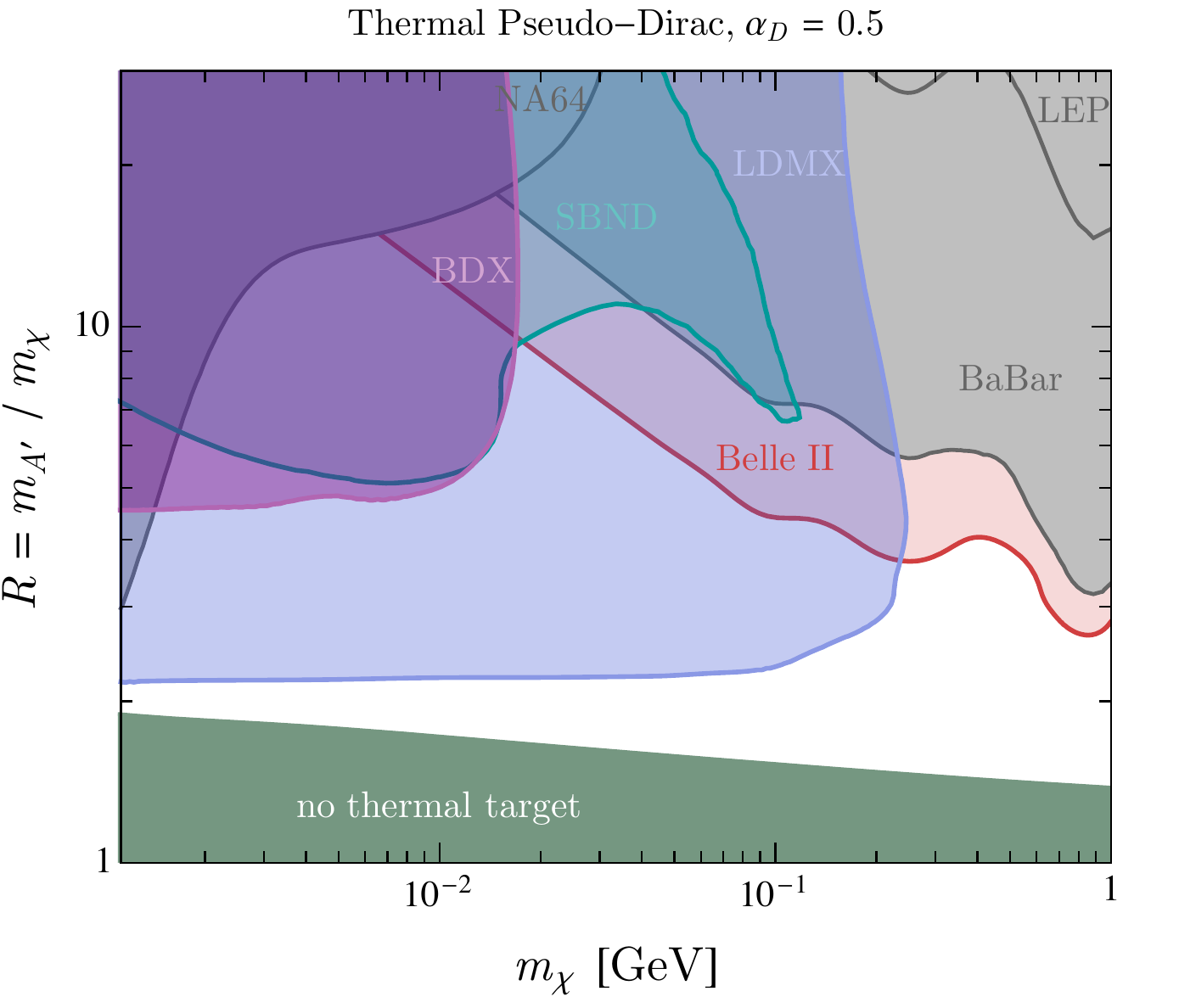}
\caption{Existing constraints (shaded gray) and projected sensitivities (color) of a representative subset of accelerator experiments to light \emph{thermal} dark matter in the $R-m_\x$ plane for fixed values of the hidden sector coupling, $\alpha_D = 0.01$ (top-left panel), $\alpha_D = 0.1$ (top-right panel), and $\alpha_D = 0.5$ (bottom panel). For each value of $R$, $m_\x$, and $\alpha_D$, the kinetic mixing parameter, $\eps$, is fixed such that the relic abundance of $\x$ agrees with the observed DM energy density. Many of the important features can be inferred directly from Fig.~\ref{fig:yR1}. We note that if the pseudo-Dirac mass splitting is sufficiently small, direct detection constraints, e.g., from CRESST-III \cite{Abdelhameed:2019hmk}, would provide complementary sensitivity on the right hand side of these plots for $m_\ch \gtrsim 300$~MeV. 
} \label{fig:yRthermal}
\end{figure}

Compared to fixed-target experiments, higher energy accelerators, such as LEP, BaBar, and Belle II, are typically most sensitive to thermal relics for larger dark photon masses. This is due to the fact that the center of mass energy for BaBar and LEP is approximately $E_\text{cm} \sim 10 \ \GeV$ and $E_\text{cm} \sim 200 \ \GeV$, respectively. Hence, for most of the parameter space that we consider, DM production proceeds through on-shell dark photons, and for $\order{1}$ $\Ap$ branching ratios to DM pairs the signal rate is independent of $\mAp$ or $m_\x$, i.e., $N_\text{sig} \propto \eps^2$. Therefore, for $2 m_\x \lesssim \mAp \ll E_\text{cm}$, the sensitivity in the $y - R$ plane continues to scale as
\be
\label{eq:yLEP}
y_\text{reach} (\text{LEP/BaBar/Belle II}) \propto \alpha_D \, R^{-4} \quad (2 m_\x \lesssim \mAp \ll E_\text{cm}),
\ee
up to much larger mass ratios, compared to lower energy experiments. The $R$ and $\alpha_D$ scaling of Eq.~(\ref{eq:yLEP}) can be seen directly in the behavior of the LEP, BaBar, and Belle II contours (shaded gray and red) of Figs.~\ref{fig:yR1} and \ref{fig:yR2}.

\subsection{The $R-m_{\ch}$ plane}

To further explore light \emph{thermal} DM, we now turn to a second two-dimensional slice of parameter space, namely the $R-m_{\ch}$ plane. In this case, for every value of $R$, $m_\x$, and $\alpha_D$, we fix the kinetic mixing parameter $\eps$ such that the relic abundance of $\x$ agrees with the observed DM energy density. This analysis requires extensive simulation and is shown for  $\alpha_D = 0.01$, $0.1$, and $0.5$ in Fig.~\ref{fig:yRthermal} for a representative subset of accelerator experiments. While we focus on pseudo-Dirac DM for concreteness, the results for other DM models such as those involving Majorana or scalar DM are  qualitatively very similar (see Fig.~\ref{fig:yR2}). In fact, for these other models, the accelerator reach to thermal targets is somewhat enhanced due to the larger couplings needed for adequate freeze-out in the early universe~\cite{Berlin:2018bsc}.

Although the main features in Fig.~\ref{fig:yRthermal} can be inferred directly from Fig.~\ref{fig:yR1}, we summarize some of the most relevant points below:
     
\begin{itemize}
\item {\it Resonant thermal relic gap} - In each panel of Fig.~\ref{fig:yRthermal}, there is a gap in sensitivity 
for $R \sim 2$.
This is due to the fact that the resonantly enhanced annihilation near freeze-out in the early universe requires much smaller couplings, as shown in the $y-R$ plane of Figs.~\ref{fig:yR1}-\ref{LBplot}. As a result, the sensitivity of any experiment to light thermal DM is inhibited in this parameter regime.
\item {\it EFT regime} - As discussed in Sec.~\ref{sec:yR}, for $R \gg 1$ and fixed DM mass, both the fixed-target reach and thermal target scale as $y \propto \text{constant}$. This saturation point occurs once DM annihilation/production is mediated through highly off-shell mediators. As a result, the sensitivity regions of Fig.~\ref{fig:yRthermal} become insensitive to $R$ for $R \gg 1$. The figures are truncated ($R \lesssim 30$) to focus on the most interesting parameter region. Extending to larger values of $R$ simply pushes further into this EFT regime, which is already excluded by existing limits. Hence, Fig.~\ref{fig:yRthermal}, along with Fig.~\ref{figEFT}, demonstrates that for $m_\x \lesssim \text{few} \times 100 \ \MeV$, a generic class of thermal DM models can be generically tested by various fixed-target experiments, independent of the dark sector coupling ($\alpha_D$) and mediator-to-DM mass ratio ($R$).
\item {\it Resonant production} - The projected sensitivity of proton fixed-target experiments approaches that of missing-momentum experiments, such as LDMX, only in special cases, where resonant mixing of the mediator with vector mesons enhances production through proton bremsstrahlung. In Fig.~\ref{fig:yRthermal}, this is apparent for SBND in the region $\mAp \sim m_\rho$.
\end{itemize}

\subsection{Comparison to Direct Detection}
Although we began by contrasting the generic validity of the EFT for direct detection with the rich dependence of accelerator-based searches on the mediator properties, we have not explicitly shown constraints from direct detection, nor the prospects from the upcoming generation of electron scattering experiments (see, e.g., Refs.~\cite{Essig:2011nj,Essig:2012yx,Angloher:2015ewa,Hochberg:2015pha,Essig:2015cda,Abramoff:2019dfb}).  For the pseudo-Dirac and scalar models considered in this paper, a relative mass splitting as small as $\order{10^{-6}}$ would forbid tree-level DM scattering in direct detection experiments, such that their sensitivity is not phenomenologically relevant.  Such mass splittings are not forbidden by an unbroken symmetry, and so there is no generic expectation that they should be small.  Because of this model-dependence, we have not shown direct detection constraints on the plots in this section.  Nonetheless, because their $y$-sensitivity is independent of $R$ in the range of interest for simple thermal DM models, it is straightforward to summarize the sensitivity of direct detection experiments for the favorable case of splittings below this level (where elastic tree-level scattering is allowed).   In this case, the strongest constraints at our $m_\x = 10$ (100) MeV benchmark points come from XENON10 electron-scattering constraints \cite{Angle:2011th,Essig:2012yx,Essig:2017kqs}, which translate to $R$-independent limit of roughly $y>2 \times 10^{-10}$ ($10^{-7}$) at the two benchmark DM masses. These constraints are not competitive with current accelerator-based limits, except in the region $R<2$.   Future experiments hope to improve on this sensitivity by four orders of magnitude or more, making them comparable in sensitivity to proposed accelerator-based experiments for models with small mass splittings (allowing elastic scattering) and velocity-independent interactions. Qualitatively similar conclusions hold for all $m_\chi \lesssim 300 \text{ MeV}$.  For DM masses between 300 MeV and a GeV, constraints on DM-nucleon scattering from CRESST III \cite{Abdelhameed:2019hmk} extend somewhat below the BaBar limits in Fig.~\ref{fig:yRthermal}.   The strongest complementarity between accelerator-based and direct detection experiments lies in the fact that the former do not rely on the assumptions of velocity-independent interactions or near-zero DM mass splittings, while the latter have excellent sensitivity to models with parametrically light mediators (in our language $R\ll 1$), which are not relevant for thermal freeze-out but can arise, for example, in DM freeze-in scenarios~\cite{Hall:2009bx}.

\section{Concluding Remarks}
\label{sec:conc}

In this paper, we have analyzed the sensitivity of fixed-target experiments to a broad class of sub-GeV thermal relic dark matter models, across a wide range of parameter space, including variations in both the mediator and dark matter mass. This allowed us to assess the reach of existing and planned electron and proton beam fixed-target experiments to light dark matter production through both on- and off-shell mediators. It has become conventional to work with the fixed slice in which dark matter can be produced by on-shell mediator decays, but our results indicate that the reach of fixed-target experiments extends well into the off-shell production regime, including the effective field theory regime where the mediator decouples. We investigated new slices of parameter space, such as the $y-R$ and $R - m_\x$ planes (where $R=\mAp/m_\x$), in order to illustrate how the reach of different experiments scales as a function of mass, in relation to the thermal target. There is a resonance for $\mAp \sim 2 m_\ch$, which renders annihilation more efficient in the early universe, and our analysis indicates that there currently exists only  a small band near this region that will remain untested for simple models coupled through the vector portal. 

\begin{acknowledgments}
We thank Richard Van de Water for helpful discussions regarding CCM. AB is supported by the James Arthur Fellowship. The work of PdN was supported in part by IBS (Project Code IBS-R018-D1) and Los Alamos National Laboratory under the LDRD program. The work of  AR is supported 
in part by NSERC, Canada. PS and NT are supported by the U.S. Department of Energy under Contract No.
684 DE-AC02-76SF00515
\end{acknowledgments}

\bibliography{DMA}

\begin{thebibliography}{10}%
\makeatletter
\providecommand \@ifxundefined [1]{%
 \ifx #1\undefined \expandafter \@firstoftwo
 \else \expandafter \@secondoftwo
\fi
}%
\providecommand \@ifnum [1]{%
 \ifnum #1\expandafter \@firstoftwo
 \else \expandafter \@secondoftwo
\fi
}%
\providecommand \enquote [1]{``#1''}%
\providecommand \bibnamefont  [1]{#1}%
\providecommand \bibfnamefont [1]{#1}%
\providecommand \citenamefont [1]{#1}%
\providecommand\href[0]{\@sanitize\@href}%
\providecommand\@href[1]{\endgroup\@@startlink{#1}\endgroup\@@href}%
\providecommand\@@href[1]{#1\@@endlink}%
\providecommand \@sanitize [0]{\begingroup\catcode`\&12\catcode`\#12\relax}%
\@ifxundefined \pdfoutput {\@firstoftwo}{%
 \@ifnum{\z@=\pdfoutput}{\@firstoftwo}{\@secondoftwo}%
}{%
 \providecommand\@@startlink[1]{\leavevmode\special{html:<a href="#1">}}%
 \providecommand\@@endlink[0]{\special{html:</a>}}%
}{%
 \providecommand\@@startlink[1]{%
  \leavevmode
  \pdfstartlink
   attr{/Border[0 0 1 ]/H/I/C[0 1 1]}%
   user{/Subtype/Link/A<</Type/Action/S/URI/URI(#1)>>}%
  \relax
 }%
 \providecommand\@@endlink[0]{\pdfendlink}%
}%
\providecommand \url  [0]{\begingroup\@sanitize \@url }%
\providecommand \@url [1]{\endgroup\@href {#1}{\urlprefix}}%
\providecommand \urlprefix [0]{URL }%
\providecommand \Eprint[0]{\href }%
\@ifxundefined \urlstyle {%
  \providecommand \doi [1]{doi:\discretionary{}{}{}#1}%
}{%
  \providecommand \doi [0]{doi:\discretionary{}{}{}\begingroup
  \urlstyle{rm}\Url }%
}%
\providecommand \doibase [0]{http://dx.doi.org/}%
\providecommand \Doi[1]{\href{\doibase#1}}%
\providecommand \bibAnnote [3]{%
  \BibitemShut{#1}%
  \begin{quotation}\noindent
    \textsc{Key:}\ #2\\\textsc{Annotation:}\ #3%
  \end{quotation}%
}%
\providecommand \bibAnnoteFile [2]{%
  \IfFileExists{#2}{\bibAnnote {#1} {#2} {\input{#2}}}{}%
}%
\providecommand \typeout [0]{\immediate \write \m@ne }%
\providecommand \selectlanguage [0]{\@gobble}%
\providecommand \bibinfo [0]{\@secondoftwo}%
\providecommand \bibfield [0]{\@secondoftwo}%
\providecommand \translation [1]{[#1]}%
\providecommand \BibitemOpen[0]{}%
\providecommand \bibitemStop [0]{}%
\providecommand \bibitemNoStop [0]{.\EOS\space}%
\providecommand \EOS [0]{\spacefactor3000\relax}%
\providecommand \BibitemShut [1]{\csname bibitem#1\endcsname}%
\bibitem{LW}%
  \BibitemOpen
  \bibfield{author}{%
  \bibinfo {author} {\bibfnamefont{B.~W.}\ \bibnamefont{Lee}}\ and\ \bibinfo
  {author} {\bibfnamefont{S.}~\bibnamefont{Weinberg}},\ }%
  \bibfield{journal}{%
  \Doi{10.1103/PhysRevLett.39.165}{\bibinfo {journal} {Phys. Rev. Lett.}}\ }%
  \textbf{\bibinfo {volume} {39}},\ \bibinfo {pages} {165} (\bibinfo {year}
  {1977}),\ \bibinfo {note} {[,183(1977)]}%
  \bibAnnoteFile{NoStop}{LW}%
\bibitem{Berlin:2017ftj}%
  \BibitemOpen
  \bibfield{author}{%
  \bibinfo {author} {\bibfnamefont{A.}~\bibnamefont{Berlin}}\ and\ \bibinfo
  {author} {\bibfnamefont{N.}~\bibnamefont{Blinov}},\ }%
  \bibfield{journal}{%
  \Doi{10.1103/PhysRevLett.120.021801}{\bibinfo {journal} {Phys. Rev. Lett.}}\
  }%
  \textbf{\bibinfo {volume} {120}},\ \bibinfo {pages} {021801} (\bibinfo {year}
  {2018}),\ \Eprint{http://arxiv.org/abs/1706.07046}{arXiv:1706.07046
  [hep-ph]}%
  \bibAnnoteFile{NoStop}{Berlin:2017ftj}%
\bibitem{Berlin:2018ztp}%
  \BibitemOpen
  \bibfield{author}{%
  \bibinfo {author} {\bibfnamefont{A.}~\bibnamefont{Berlin}}\ and\ \bibinfo
  {author} {\bibfnamefont{N.}~\bibnamefont{Blinov}},\ }%
  \bibfield{journal}{%
  \Doi{10.1103/PhysRevD.99.095030}{\bibinfo {journal} {Phys. Rev.}}\ }%
  \textbf{\bibinfo {volume} {D99}},\ \bibinfo {pages} {095030} (\bibinfo {year}
  {2019}),\ \Eprint{http://arxiv.org/abs/1807.04282}{arXiv:1807.04282
  [hep-ph]}%
  \bibAnnoteFile{NoStop}{Berlin:2018ztp}%
\bibitem{Boehm:2003hm}%
  \BibitemOpen
  \bibfield{author}{%
  \bibinfo {author} {\bibfnamefont{C.}~\bibnamefont{Boehm}}\ and\ \bibinfo
  {author} {\bibfnamefont{P.}~\bibnamefont{Fayet}},\ }%
  \bibfield{journal}{%
  \Doi{10.1016/j.nuclphysb.2004.01.015}{\bibinfo {journal} {Nucl.Phys.}}\ }%
  \textbf{\bibinfo {volume} {B683}},\ \bibinfo {pages} {219} (\bibinfo {year}
  {2004}),\ \Eprint{http://arxiv.org/abs/hep-ph/0305261}{arXiv:hep-ph/0305261
  [hep-ph]}%
  \bibAnnoteFile{NoStop}{Boehm:2003hm}%
\bibitem{DS16}%
  \BibitemOpen
  \bibfield{author}{%
  \bibinfo {author} {\bibfnamefont{J.}~\bibnamefont{Alexander}} \emph{et~al.}\
  }%
  (\bibinfo {year} {2016})\
  \Eprint{http://arxiv.org/abs/1608.08632}{arXiv:1608.08632 [hep-ph]},\
  \url{http://lss.fnal.gov/archive/2016/conf/fermilab-conf-16-421.pdf}%
  \bibAnnoteFile{NoStop}{DS16}%
\bibitem{CV17}%
  \BibitemOpen
  \bibfield{author}{%
  \bibinfo {author} {\bibfnamefont{M.}~\bibnamefont{Battaglieri}}
  \emph{et~al.},\ }%
  in\ \emph{\bibinfo {booktitle} {{U.S. Cosmic Visions: New Ideas in Dark
  Matter College Park, MD, USA, March 23-25, 2017}}}\ (\bibinfo {year} {2017})\
  \Eprint{http://arxiv.org/abs/1707.04591}{arXiv:1707.04591 [hep-ph]},\
  \url{http://lss.fnal.gov/archive/2017/conf/fermilab-conf-17-282-ae-ppd-t.pdf%
}%
  \bibAnnoteFile{NoStop}{CV17}%
\bibitem{PBC}%
  \BibitemOpen
  \bibfield{author}{%
  \bibinfo {author} {\bibfnamefont{J.}~\bibnamefont{Beacham}} \emph{et~al.}}%
   (\bibinfo {year} {2019}),\
  \Eprint{http://arxiv.org/abs/1901.09966}{arXiv:1901.09966 [hep-ex]}%
  \bibAnnoteFile{NoStop}{PBC}%
\bibitem{Essig:2011nj}%
  \BibitemOpen
  \bibfield{author}{%
  \bibinfo {author} {\bibfnamefont{R.}~\bibnamefont{Essig}}, \bibinfo {author}
  {\bibfnamefont{J.}~\bibnamefont{Mardon}},\ and\ \bibinfo {author}
  {\bibfnamefont{T.}~\bibnamefont{Volansky}},\ }%
  \bibfield{journal}{%
  \Doi{10.1103/PhysRevD.85.076007}{\bibinfo {journal} {Phys. Rev.}}\ }%
  \textbf{\bibinfo {volume} {D85}},\ \bibinfo {pages} {076007} (\bibinfo {year}
  {2012}),\ \Eprint{http://arxiv.org/abs/1108.5383}{arXiv:1108.5383 [hep-ph]}%
  \bibAnnoteFile{NoStop}{Essig:2011nj}%
\bibitem{Essig:2012yx}%
  \BibitemOpen
  \bibfield{author}{%
  \bibinfo {author} {\bibfnamefont{R.}~\bibnamefont{Essig}}, \bibinfo {author}
  {\bibfnamefont{A.}~\bibnamefont{Manalaysay}}, \bibinfo {author}
  {\bibfnamefont{J.}~\bibnamefont{Mardon}}, \bibinfo {author}
  {\bibfnamefont{P.}~\bibnamefont{Sorensen}},\ and\ \bibinfo {author}
  {\bibfnamefont{T.}~\bibnamefont{Volansky}},\ }%
  \bibfield{journal}{%
  \Doi{10.1103/PhysRevLett.109.021301}{\bibinfo {journal} {Phys. Rev. Lett.}}\
  }%
  \textbf{\bibinfo {volume} {109}},\ \bibinfo {pages} {021301} (\bibinfo {year}
  {2012}),\ \Eprint{http://arxiv.org/abs/1206.2644}{arXiv:1206.2644
  [astro-ph.CO]}%
  \bibAnnoteFile{NoStop}{Essig:2012yx}%
\bibitem{Angloher:2015ewa}%
  \BibitemOpen
  \bibfield{author}{%
  \bibinfo {author} {\bibfnamefont{G.}~\bibnamefont{Angloher}} \emph{et~al.}
  (\bibinfo {collaboration} {CRESST}),\ }%
  \bibfield{journal}{%
  \Doi{10.1140/epjc/s10052-016-3877-3}{\bibinfo {journal} {Eur. Phys. J.}}\ }%
  \textbf{\bibinfo {volume} {C76}},\ \bibinfo {pages} {25} (\bibinfo {year}
  {2016}),\ \Eprint{http://arxiv.org/abs/1509.01515}{arXiv:1509.01515
  [astro-ph.CO]}%
  \bibAnnoteFile{NoStop}{Angloher:2015ewa}%
\bibitem{Hochberg:2015pha}%
  \BibitemOpen
  \bibfield{author}{%
  \bibinfo {author} {\bibfnamefont{Y.}~\bibnamefont{Hochberg}}, \bibinfo
  {author} {\bibfnamefont{Y.}~\bibnamefont{Zhao}},\ and\ \bibinfo {author}
  {\bibfnamefont{K.~M.}\ \bibnamefont{Zurek}},\ }%
  \bibfield{journal}{%
  \Doi{10.1103/PhysRevLett.116.011301}{\bibinfo {journal} {Phys. Rev. Lett.}}\
  }%
  \textbf{\bibinfo {volume} {116}},\ \bibinfo {pages} {011301} (\bibinfo {year}
  {2016}),\ \Eprint{http://arxiv.org/abs/1504.07237}{arXiv:1504.07237
  [hep-ph]}%
  \bibAnnoteFile{NoStop}{Hochberg:2015pha}%
\bibitem{Essig:2015cda}%
  \BibitemOpen
  \bibfield{author}{%
  \bibinfo {author} {\bibfnamefont{R.}~\bibnamefont{Essig}}, \bibinfo {author}
  {\bibfnamefont{M.}~\bibnamefont{Fernandez-Serra}}, \bibinfo {author}
  {\bibfnamefont{J.}~\bibnamefont{Mardon}}, \bibinfo {author}
  {\bibfnamefont{A.}~\bibnamefont{Soto}}, \bibinfo {author}
  {\bibfnamefont{T.}~\bibnamefont{Volansky}},\ and\ \bibinfo {author}
  {\bibfnamefont{T.-T.}\ \bibnamefont{Yu}},\ }%
  \bibfield{journal}{%
  \Doi{10.1007/JHEP05(2016)046}{\bibinfo {journal} {JHEP}}\ }%
  \textbf{\bibinfo {volume} {05}},\ \bibinfo {pages} {046} (\bibinfo {year}
  {2016}),\ \Eprint{http://arxiv.org/abs/1509.01598}{arXiv:1509.01598
  [hep-ph]}%
  \bibAnnoteFile{NoStop}{Essig:2015cda}%
\bibitem{Abdelhameed:2019hmk}%
  \BibitemOpen
  \bibfield{author}{%
  \bibinfo {author} {\bibfnamefont{A.~H.}\ \bibnamefont{Abdelhameed}}
  \emph{et~al.} (\bibinfo {collaboration} {CRESST}),\ }%
  \bibfield{journal}{%
  \Doi{10.1103/PhysRevD.100.102002}{\bibinfo {journal} {Phys. Rev.}}\ }%
  \textbf{\bibinfo {volume} {D100}},\ \bibinfo {pages} {102002} (\bibinfo
  {year} {2019}),\ \Eprint{http://arxiv.org/abs/1904.00498}{arXiv:1904.00498
  [astro-ph.CO]}%
  \bibAnnoteFile{NoStop}{Abdelhameed:2019hmk}%
\bibitem{Abramoff:2019dfb}%
  \BibitemOpen
  \bibfield{author}{%
  \bibinfo {author} {\bibfnamefont{O.}~\bibnamefont{Abramoff}} \emph{et~al.}
  (\bibinfo {collaboration} {SENSEI}),\ }%
  \bibfield{journal}{%
  \Doi{10.1103/PhysRevLett.122.161801}{\bibinfo {journal} {Phys. Rev. Lett.}}\
  }%
  \textbf{\bibinfo {volume} {122}},\ \bibinfo {pages} {161801} (\bibinfo {year}
  {2019}),\ \Eprint{http://arxiv.org/abs/1901.10478}{arXiv:1901.10478
  [hep-ex]}%
  \bibAnnoteFile{NoStop}{Abramoff:2019dfb}%
\bibitem{Bjorken:2009mm}%
  \BibitemOpen
  \bibfield{author}{%
  \bibinfo {author} {\bibfnamefont{J.~D.}\ \bibnamefont{Bjorken}}, \bibinfo
  {author} {\bibfnamefont{R.}~\bibnamefont{Essig}}, \bibinfo {author}
  {\bibfnamefont{P.}~\bibnamefont{Schuster}},\ and\ \bibinfo {author}
  {\bibfnamefont{N.}~\bibnamefont{Toro}},\ }%
  \bibfield{journal}{%
  \Doi{10.1103/PhysRevD.80.075018}{\bibinfo {journal} {Phys.Rev.}}\ }%
  \textbf{\bibinfo {volume} {D80}},\ \bibinfo {pages} {075018} (\bibinfo {year}
  {2009}),\ \Eprint{http://arxiv.org/abs/0906.0580}{arXiv:0906.0580 [hep-ph]}%
  \bibAnnoteFile{NoStop}{Bjorken:2009mm}%
\bibitem{Izaguirre:2013uxa}%
  \BibitemOpen
  \bibfield{author}{%
  \bibinfo {author} {\bibfnamefont{E.}~\bibnamefont{Izaguirre}}, \bibinfo
  {author} {\bibfnamefont{G.}~\bibnamefont{Krnjaic}}, \bibinfo {author}
  {\bibfnamefont{P.}~\bibnamefont{Schuster}},\ and\ \bibinfo {author}
  {\bibfnamefont{N.}~\bibnamefont{Toro}},\ }%
  \bibfield{journal}{%
  \Doi{10.1103/PhysRevD.88.114015}{\bibinfo {journal} {Phys.Rev.}}\ }%
  \textbf{\bibinfo {volume} {D88}},\ \bibinfo {pages} {114015} (\bibinfo {year}
  {2013}),\ \Eprint{http://arxiv.org/abs/1307.6554}{arXiv:1307.6554 [hep-ph]}%
  \bibAnnoteFile{NoStop}{Izaguirre:2013uxa}%
\bibitem{Diamond:2013oda}%
  \BibitemOpen
  \bibfield{author}{%
  \bibinfo {author} {\bibfnamefont{M.~D.}\ \bibnamefont{Diamond}}\ and\
  \bibinfo {author} {\bibfnamefont{P.}~\bibnamefont{Schuster}},\ }%
  \bibfield{journal}{%
  \Doi{10.1103/PhysRevLett.111.221803}{\bibinfo {journal} {Phys.Rev.Lett.}}\ }%
  \textbf{\bibinfo {volume} {111}},\ \bibinfo {pages} {221803} (\bibinfo {year}
  {2013}),\ \Eprint{http://arxiv.org/abs/1307.6861}{arXiv:1307.6861 [hep-ph]}%
  \bibAnnoteFile{NoStop}{Diamond:2013oda}%
\bibitem{Izaguirre:2014dua}%
  \BibitemOpen
  \bibfield{author}{%
  \bibinfo {author} {\bibfnamefont{E.}~\bibnamefont{Izaguirre}}, \bibinfo
  {author} {\bibfnamefont{G.}~\bibnamefont{Krnjaic}}, \bibinfo {author}
  {\bibfnamefont{P.}~\bibnamefont{Schuster}},\ and\ \bibinfo {author}
  {\bibfnamefont{N.}~\bibnamefont{Toro}}}%
   (\bibinfo {year} {2014}),\
  \Eprint{http://arxiv.org/abs/1403.6826}{arXiv:1403.6826 [hep-ph]}%
  \bibAnnoteFile{NoStop}{Izaguirre:2014dua}%
\bibitem{Batell:2014mga}%
  \BibitemOpen
  \bibfield{author}{%
  \bibinfo {author} {\bibfnamefont{B.}~\bibnamefont{Batell}}, \bibinfo {author}
  {\bibfnamefont{R.}~\bibnamefont{Essig}},\ and\ \bibinfo {author}
  {\bibfnamefont{Z.}~\bibnamefont{Surujon}},\ }%
  \bibfield{journal}{%
  \Doi{10.1103/PhysRevLett.113.171802}{\bibinfo {journal} {Phys.Rev.Lett.}}\ }%
  \textbf{\bibinfo {volume} {113}},\ \bibinfo {pages} {171802} (\bibinfo {year}
  {2014}),\ \Eprint{http://arxiv.org/abs/1406.2698}{arXiv:1406.2698 [hep-ph]}%
  \bibAnnoteFile{NoStop}{Batell:2014mga}%
\bibitem{Lees:2017lec}%
  \BibitemOpen
  \bibfield{author}{%
  \bibinfo {author} {\bibfnamefont{J.~P.}\ \bibnamefont{Lees}} \emph{et~al.}
  (\bibinfo {collaboration} {BaBar}),\ }%
  \bibfield{journal}{%
  \Doi{10.1103/PhysRevLett.119.131804}{\bibinfo {journal} {Phys. Rev. Lett.}}\
  }%
  \textbf{\bibinfo {volume} {119}},\ \bibinfo {pages} {131804} (\bibinfo {year}
  {2017}),\ \Eprint{http://arxiv.org/abs/1702.03327}{arXiv:1702.03327
  [hep-ex]}%
  \bibAnnoteFile{NoStop}{Lees:2017lec}%
\bibitem{Berlin:2018bsc}%
  \BibitemOpen
  \bibfield{author}{%
  \bibinfo {author} {\bibfnamefont{A.}~\bibnamefont{Berlin}}, \bibinfo {author}
  {\bibfnamefont{N.}~\bibnamefont{Blinov}}, \bibinfo {author}
  {\bibfnamefont{G.}~\bibnamefont{Krnjaic}}, \bibinfo {author}
  {\bibfnamefont{P.}~\bibnamefont{Schuster}},\ and\ \bibinfo {author}
  {\bibfnamefont{N.}~\bibnamefont{Toro}},\ }%
  \bibfield{journal}{%
  \Doi{10.1103/PhysRevD.99.075001}{\bibinfo {journal} {Phys. Rev.}}\ }%
  \textbf{\bibinfo {volume} {D99}},\ \bibinfo {pages} {075001} (\bibinfo {year}
  {2019}),\ \Eprint{http://arxiv.org/abs/1807.01730}{arXiv:1807.01730
  [hep-ph]}%
  \bibAnnoteFile{NoStop}{Berlin:2018bsc}%
\bibitem{NA64:2019imj}%
  \BibitemOpen
  \bibfield{author}{%
  \bibinfo {author} {\bibfnamefont{D.}~\bibnamefont{Banerjee}} \emph{et~al.},\
  }%
  \bibfield{journal}{%
  \Doi{10.1103/PhysRevLett.123.121801}{\bibinfo {journal} {Phys. Rev. Lett.}}\
  }%
  \textbf{\bibinfo {volume} {123}},\ \bibinfo {pages} {121801} (\bibinfo {year}
  {2019}),\ \Eprint{http://arxiv.org/abs/1906.00176}{arXiv:1906.00176
  [hep-ex]}%
  \bibAnnoteFile{NoStop}{NA64:2019imj}%
\bibitem{Batell:2009di}%
  \BibitemOpen
  \bibfield{author}{%
  \bibinfo {author} {\bibfnamefont{B.}~\bibnamefont{Batell}}, \bibinfo {author}
  {\bibfnamefont{M.}~\bibnamefont{Pospelov}},\ and\ \bibinfo {author}
  {\bibfnamefont{A.}~\bibnamefont{Ritz}},\ }%
  \bibfield{journal}{%
  \Doi{10.1103/PhysRevD.80.095024}{\bibinfo {journal} {Phys.Rev.}}\ }%
  \textbf{\bibinfo {volume} {D80}},\ \bibinfo {pages} {095024} (\bibinfo {year}
  {2009}),\ \Eprint{http://arxiv.org/abs/0906.5614}{arXiv:0906.5614 [hep-ph]}%
  \bibAnnoteFile{NoStop}{Batell:2009di}%
\bibitem{deNiverville:2011it}%
  \BibitemOpen
  \bibfield{author}{%
  \bibinfo {author} {\bibfnamefont{P.}~\bibnamefont{deNiverville}}, \bibinfo
  {author} {\bibfnamefont{M.}~\bibnamefont{Pospelov}},\ and\ \bibinfo {author}
  {\bibfnamefont{A.}~\bibnamefont{Ritz}},\ }%
  \bibfield{journal}{%
  \Doi{10.1103/PhysRevD.84.075020}{\bibinfo {journal} {Phys.Rev.}}\ }%
  \textbf{\bibinfo {volume} {D84}},\ \bibinfo {pages} {075020} (\bibinfo {year}
  {2011}),\ \Eprint{http://arxiv.org/abs/1107.4580}{arXiv:1107.4580 [hep-ph]}%
  \bibAnnoteFile{NoStop}{deNiverville:2011it}%
\bibitem{deNiverville:2012ij}%
  \BibitemOpen
  \bibfield{author}{%
  \bibinfo {author} {\bibfnamefont{P.}~\bibnamefont{deNiverville}}, \bibinfo
  {author} {\bibfnamefont{D.}~\bibnamefont{McKeen}},\ and\ \bibinfo {author}
  {\bibfnamefont{A.}~\bibnamefont{Ritz}},\ }%
  \bibfield{journal}{%
  \Doi{10.1103/PhysRevD.86.035022}{\bibinfo {journal} {Phys.Rev.}}\ }%
  \textbf{\bibinfo {volume} {D86}},\ \bibinfo {pages} {035022} (\bibinfo {year}
  {2012}),\ \Eprint{http://arxiv.org/abs/1205.3499}{arXiv:1205.3499 [hep-ph]}%
  \bibAnnoteFile{NoStop}{deNiverville:2012ij}%
\bibitem{Kahn:2014sra}%
  \BibitemOpen
  \bibfield{author}{%
  \bibinfo {author} {\bibfnamefont{Y.}~\bibnamefont{Kahn}}, \bibinfo {author}
  {\bibfnamefont{G.}~\bibnamefont{Krnjaic}}, \bibinfo {author}
  {\bibfnamefont{J.}~\bibnamefont{Thaler}},\ and\ \bibinfo {author}
  {\bibfnamefont{M.}~\bibnamefont{Toups}}}%
   (\bibinfo {year} {2014}),\
  \Eprint{http://arxiv.org/abs/1411.1055}{arXiv:1411.1055 [hep-ph]}%
  \bibAnnoteFile{NoStop}{Kahn:2014sra}%
\bibitem{Adams:2013qkq}%
  \BibitemOpen
  \bibfield{author}{%
  \bibinfo {author} {\bibfnamefont{C.}~\bibnamefont{Adams}} \emph{et~al.}
  (\bibinfo {collaboration} {LBNE})\ }%
  (\bibinfo {year} {2013})\
  \Eprint{http://arxiv.org/abs/1307.7335}{arXiv:1307.7335 [hep-ex]},\
  \url{http://www.osti.gov/scitech/biblio/1128102}%
  \bibAnnoteFile{NoStop}{Adams:2013qkq}%
\bibitem{Soper:2014ska}%
  \BibitemOpen
  \bibfield{author}{%
  \bibinfo {author} {\bibfnamefont{D.~E.}\ \bibnamefont{Soper}}, \bibinfo
  {author} {\bibfnamefont{M.}~\bibnamefont{Spannowsky}}, \bibinfo {author}
  {\bibfnamefont{C.~J.}\ \bibnamefont{Wallace}},\ and\ \bibinfo {author}
  {\bibfnamefont{T.~M.~P.}\ \bibnamefont{Tait}},\ }%
  \bibfield{journal}{%
  \Doi{10.1103/PhysRevD.90.115005}{\bibinfo {journal} {Phys. Rev.}}\ }%
  \textbf{\bibinfo {volume} {D90}},\ \bibinfo {pages} {115005} (\bibinfo {year}
  {2014}),\ \Eprint{http://arxiv.org/abs/1407.2623}{arXiv:1407.2623 [hep-ph]}%
  \bibAnnoteFile{NoStop}{Soper:2014ska}%
\bibitem{Dobrescu:2014ita}%
  \BibitemOpen
  \bibfield{author}{%
  \bibinfo {author} {\bibfnamefont{B.~A.}\ \bibnamefont{Dobrescu}}\ and\
  \bibinfo {author} {\bibfnamefont{C.}~\bibnamefont{Frugiuele}},\ }%
  \bibfield{journal}{%
  \Doi{10.1007/JHEP02(2015)019}{\bibinfo {journal} {JHEP}}\ }%
  \textbf{\bibinfo {volume} {02}},\ \bibinfo {pages} {019} (\bibinfo {year}
  {2015}),\ \Eprint{http://arxiv.org/abs/1410.1566}{arXiv:1410.1566 [hep-ph]}%
  \bibAnnoteFile{NoStop}{Dobrescu:2014ita}%
\bibitem{Coloma:2015pih}%
  \BibitemOpen
  \bibfield{author}{%
  \bibinfo {author} {\bibfnamefont{P.}~\bibnamefont{Coloma}}, \bibinfo {author}
  {\bibfnamefont{B.~A.}\ \bibnamefont{Dobrescu}}, \bibinfo {author}
  {\bibfnamefont{C.}~\bibnamefont{Frugiuele}},\ and\ \bibinfo {author}
  {\bibfnamefont{R.}~\bibnamefont{Harnik}},\ }%
  \bibfield{journal}{%
  \Doi{10.1007/JHEP04(2016)047}{\bibinfo {journal} {JHEP}}\ }%
  \textbf{\bibinfo {volume} {04}},\ \bibinfo {pages} {047} (\bibinfo {year}
  {2016}),\ \Eprint{http://arxiv.org/abs/1512.03852}{arXiv:1512.03852
  [hep-ph]}%
  \bibAnnoteFile{NoStop}{Coloma:2015pih}%
\bibitem{dNCPR}%
  \BibitemOpen
  \bibfield{author}{%
  \bibinfo {author} {\bibfnamefont{P.}~\bibnamefont{deNiverville}}, \bibinfo
  {author} {\bibfnamefont{C.-Y.}\ \bibnamefont{Chen}}, \bibinfo {author}
  {\bibfnamefont{M.}~\bibnamefont{Pospelov}},\ and\ \bibinfo {author}
  {\bibfnamefont{A.}~\bibnamefont{Ritz}},\ }%
  \bibfield{journal}{%
  \Doi{10.1103/PhysRevD.95.035006}{\bibinfo {journal} {Phys. Rev.}}\ }%
  \textbf{\bibinfo {volume} {D95}},\ \bibinfo {pages} {035006} (\bibinfo {year}
  {2017}),\ \Eprint{http://arxiv.org/abs/1609.01770}{arXiv:1609.01770
  [hep-ph]}%
  \bibAnnoteFile{NoStop}{dNCPR}%
\bibitem{MB1}%
  \BibitemOpen
  \bibfield{author}{%
  \bibinfo {author} {\bibfnamefont{A.~A.}\ \bibnamefont{Aguilar-Arevalo}}
  \emph{et~al.} (\bibinfo {collaboration} {MiniBooNE}),\ }%
  \bibfield{journal}{%
  \Doi{10.1103/PhysRevLett.118.221803}{\bibinfo {journal} {Phys. Rev. Lett.}}\
  }%
  \textbf{\bibinfo {volume} {118}},\ \bibinfo {pages} {221803} (\bibinfo {year}
  {2017}),\ \Eprint{http://arxiv.org/abs/1702.02688}{arXiv:1702.02688
  [hep-ex]}%
  \bibAnnoteFile{NoStop}{MB1}%
\bibitem{MB2}%
  \BibitemOpen
  \bibfield{author}{%
  \bibinfo {author} {\bibfnamefont{A.~A.}\ \bibnamefont{Aguilar-Arevalo}}
  \emph{et~al.} (\bibinfo {collaboration} {MiniBooNE DM}),\ }%
  \bibfield{journal}{%
  \Doi{10.1103/PhysRevD.98.112004}{\bibinfo {journal} {Phys. Rev.}}\ }%
  \textbf{\bibinfo {volume} {D98}},\ \bibinfo {pages} {112004} (\bibinfo {year}
  {2018}),\ \Eprint{http://arxiv.org/abs/1807.06137}{arXiv:1807.06137
  [hep-ex]}%
  \bibAnnoteFile{NoStop}{MB2}%
\bibitem{Alpigiani:2018fgd}%
  \BibitemOpen
  \bibfield{author}{%
  \bibinfo {author} {\bibfnamefont{C.}~\bibnamefont{Alpigiani}} \emph{et~al.}
  (\bibinfo {collaboration} {MATHUSLA})}%
   (\bibinfo {year} {2018}),\
  \Eprint{http://arxiv.org/abs/1811.00927}{arXiv:1811.00927 [physics.ins-det]}%
  \bibAnnoteFile{NoStop}{Alpigiani:2018fgd}%
\bibitem{Ariga:2018pin}%
  \BibitemOpen
  \bibfield{author}{%
  \bibinfo {author} {\bibfnamefont{A.}~\bibnamefont{Ariga}} \emph{et~al.}
  (\bibinfo {collaboration} {FASER})}%
   (\bibinfo {year} {2018}),\
  \Eprint{http://arxiv.org/abs/1812.09139}{arXiv:1812.09139 [physics.ins-det]}%
  \bibAnnoteFile{NoStop}{Ariga:2018pin}%
\bibitem{Hewett:2012ns}%
  \BibitemOpen
  \bibfield{author}{%
  \bibinfo {author} {\bibfnamefont{J.}~\bibnamefont{Hewett}}, \bibinfo {author}
  {\bibfnamefont{H.}~\bibnamefont{Weerts}}, \bibinfo {author}
  {\bibfnamefont{R.}~\bibnamefont{Brock}}, \bibinfo {author}
  {\bibfnamefont{J.}~\bibnamefont{Butler}}, \bibinfo {author}
  {\bibfnamefont{B.}~\bibnamefont{Casey}}, \emph{et~al.}}%
   (\bibinfo {year} {2012}),\
  \Eprint{http://arxiv.org/abs/1205.2671}{arXiv:1205.2671 [hep-ex]}%
  \bibAnnoteFile{NoStop}{Hewett:2012ns}%
\bibitem{Kronfeld:2013uoa}%
  \BibitemOpen
  \bibfield{author}{%
  \bibinfo {author} {\bibfnamefont{A.~S.}\ \bibnamefont{Kronfeld}}, \bibinfo
  {author} {\bibfnamefont{R.~S.}\ \bibnamefont{Tschirhart}}, \bibinfo {author}
  {\bibfnamefont{U.}~\bibnamefont{Al-Binni}}, \bibinfo {author}
  {\bibfnamefont{W.}~\bibnamefont{Altmannshofer}}, \bibinfo {author}
  {\bibfnamefont{C.}~\bibnamefont{Ankenbrandt}}, \emph{et~al.}}%
   (\bibinfo {year} {2013}),\
  \Eprint{http://arxiv.org/abs/1306.5009}{arXiv:1306.5009 [hep-ex]}%
  \bibAnnoteFile{NoStop}{Kronfeld:2013uoa}%
\bibitem{Essig:2013lka}%
  \BibitemOpen
  \bibfield{author}{%
  \bibinfo {author} {\bibfnamefont{R.}~\bibnamefont{Essig}}, \bibinfo {author}
  {\bibfnamefont{J.~A.}\ \bibnamefont{Jaros}}, \bibinfo {author}
  {\bibfnamefont{W.}~\bibnamefont{Wester}}, \bibinfo {author}
  {\bibfnamefont{P.~H.}\ \bibnamefont{Adrian}}, \bibinfo {author}
  {\bibfnamefont{S.}~\bibnamefont{Andreas}}, \emph{et~al.}}%
   (\bibinfo {year} {2013}),\
  \Eprint{http://arxiv.org/abs/1311.0029}{arXiv:1311.0029 [hep-ph]}%
  \bibAnnoteFile{NoStop}{Essig:2013lka}%
\bibitem{pospelov2008}%
  \BibitemOpen
  \bibfield{author}{%
  \bibinfo {author} {\bibfnamefont{M.}~\bibnamefont{Pospelov}},\ }%
  \bibfield{journal}{%
  \Doi{10.1103/PhysRevD.80.095002}{\bibinfo {journal} {Phys.Rev.}}\ }%
  \textbf{\bibinfo {volume} {D80}},\ \bibinfo {pages} {095002} (\bibinfo {year}
  {2009}),\ \Eprint{http://arxiv.org/abs/0811.1030}{arXiv:0811.1030 [hep-ph]}%
  \bibAnnoteFile{NoStop}{pospelov2008}%
\bibitem{Batell:2009yf}%
  \BibitemOpen
  \bibfield{author}{%
  \bibinfo {author} {\bibfnamefont{B.}~\bibnamefont{Batell}}, \bibinfo {author}
  {\bibfnamefont{M.}~\bibnamefont{Pospelov}},\ and\ \bibinfo {author}
  {\bibfnamefont{A.}~\bibnamefont{Ritz}},\ }%
  \bibfield{journal}{%
  \Doi{10.1103/PhysRevD.79.115008}{\bibinfo {journal} {Phys.Rev.}}\ }%
  \textbf{\bibinfo {volume} {D79}},\ \bibinfo {pages} {115008} (\bibinfo {year}
  {2009}),\ \Eprint{http://arxiv.org/abs/0903.0363}{arXiv:0903.0363 [hep-ph]}%
  \bibAnnoteFile{NoStop}{Batell:2009yf}%
\bibitem{Essig:2009nc}%
  \BibitemOpen
  \bibfield{author}{%
  \bibinfo {author} {\bibfnamefont{R.}~\bibnamefont{Essig}}, \bibinfo {author}
  {\bibfnamefont{P.}~\bibnamefont{Schuster}},\ and\ \bibinfo {author}
  {\bibfnamefont{N.}~\bibnamefont{Toro}},\ }%
  \bibfield{journal}{%
  \Doi{10.1103/PhysRevD.80.015003}{\bibinfo {journal} {Phys.Rev.}}\ }%
  \textbf{\bibinfo {volume} {D80}},\ \bibinfo {pages} {015003} (\bibinfo {year}
  {2009}),\ \Eprint{http://arxiv.org/abs/0903.3941}{arXiv:0903.3941 [hep-ph]}%
  \bibAnnoteFile{NoStop}{Essig:2009nc}%
\bibitem{Reece:2009un}%
  \BibitemOpen
  \bibfield{author}{%
  \bibinfo {author} {\bibfnamefont{M.}~\bibnamefont{Reece}}\ and\ \bibinfo
  {author} {\bibfnamefont{L.-T.}\ \bibnamefont{Wang}},\ }%
  \bibfield{journal}{%
  \Doi{10.1088/1126-6708/2009/07/051}{\bibinfo {journal} {JHEP}}\ }%
  \textbf{\bibinfo {volume} {0907}},\ \bibinfo {pages} {051} (\bibinfo {year}
  {2009}),\ \Eprint{http://arxiv.org/abs/0904.1743}{arXiv:0904.1743 [hep-ph]}%
  \bibAnnoteFile{NoStop}{Reece:2009un}%
\bibitem{Freytsis:2009bh}%
  \BibitemOpen
  \bibfield{author}{%
  \bibinfo {author} {\bibfnamefont{M.}~\bibnamefont{Freytsis}}, \bibinfo
  {author} {\bibfnamefont{G.}~\bibnamefont{Ovanesyan}},\ and\ \bibinfo {author}
  {\bibfnamefont{J.}~\bibnamefont{Thaler}},\ }%
  \bibfield{journal}{%
  \Doi{10.1007/JHEP01(2010)111}{\bibinfo {journal} {JHEP}}\ }%
  \textbf{\bibinfo {volume} {1001}},\ \bibinfo {pages} {111} (\bibinfo {year}
  {2010}),\ \Eprint{http://arxiv.org/abs/0909.2862}{arXiv:0909.2862 [hep-ph]}%
  \bibAnnoteFile{NoStop}{Freytsis:2009bh}%
\bibitem{Batell:2009jf}%
  \BibitemOpen
  \bibfield{author}{%
  \bibinfo {author} {\bibfnamefont{B.}~\bibnamefont{Batell}}, \bibinfo {author}
  {\bibfnamefont{M.}~\bibnamefont{Pospelov}},\ and\ \bibinfo {author}
  {\bibfnamefont{A.}~\bibnamefont{Ritz}},\ }%
  \bibfield{journal}{%
  \Doi{10.1103/PhysRevD.83.054005}{\bibinfo {journal} {Phys.Rev.}}\ }%
  \textbf{\bibinfo {volume} {D83}},\ \bibinfo {pages} {054005} (\bibinfo {year}
  {2011}),\ \Eprint{http://arxiv.org/abs/0911.4938}{arXiv:0911.4938 [hep-ph]}%
  \bibAnnoteFile{NoStop}{Batell:2009jf}%
\bibitem{Freytsis:2009ct}%
  \BibitemOpen
  \bibfield{author}{%
  \bibinfo {author} {\bibfnamefont{M.}~\bibnamefont{Freytsis}}, \bibinfo
  {author} {\bibfnamefont{Z.}~\bibnamefont{Ligeti}},\ and\ \bibinfo {author}
  {\bibfnamefont{J.}~\bibnamefont{Thaler}},\ }%
  \bibfield{journal}{%
  \Doi{10.1103/PhysRevD.81.034001}{\bibinfo {journal} {Phys.Rev.}}\ }%
  \textbf{\bibinfo {volume} {D81}},\ \bibinfo {pages} {034001} (\bibinfo {year}
  {2010}),\ \Eprint{http://arxiv.org/abs/0911.5355}{arXiv:0911.5355 [hep-ph]}%
  \bibAnnoteFile{NoStop}{Freytsis:2009ct}%
\bibitem{Essig:2010xa}%
  \BibitemOpen
  \bibfield{author}{%
  \bibinfo {author} {\bibfnamefont{R.}~\bibnamefont{Essig}}, \bibinfo {author}
  {\bibfnamefont{P.}~\bibnamefont{Schuster}}, \bibinfo {author}
  {\bibfnamefont{N.}~\bibnamefont{Toro}},\ and\ \bibinfo {author}
  {\bibfnamefont{B.}~\bibnamefont{Wojtsekhowski}},\ }%
  \bibfield{journal}{%
  \Doi{10.1007/JHEP02(2011)009}{\bibinfo {journal} {JHEP}}\ }%
  \textbf{\bibinfo {volume} {1102}},\ \bibinfo {pages} {009} (\bibinfo {year}
  {2011}),\ \Eprint{http://arxiv.org/abs/1001.2557}{arXiv:1001.2557 [hep-ph]}%
  \bibAnnoteFile{NoStop}{Essig:2010xa}%
\bibitem{Essig:2010gu}%
  \BibitemOpen
  \bibfield{author}{%
  \bibinfo {author} {\bibfnamefont{R.}~\bibnamefont{Essig}}, \bibinfo {author}
  {\bibfnamefont{R.}~\bibnamefont{Harnik}}, \bibinfo {author}
  {\bibfnamefont{J.}~\bibnamefont{Kaplan}},\ and\ \bibinfo {author}
  {\bibfnamefont{N.}~\bibnamefont{Toro}},\ }%
  \bibfield{journal}{%
  \Doi{10.1103/PhysRevD.82.113008}{\bibinfo {journal} {Phys.Rev.}}\ }%
  \textbf{\bibinfo {volume} {D82}},\ \bibinfo {pages} {113008} (\bibinfo {year}
  {2010}),\ \Eprint{http://arxiv.org/abs/1008.0636}{arXiv:1008.0636 [hep-ph]}%
  \bibAnnoteFile{NoStop}{Essig:2010gu}%
\bibitem{McDonald:2010fe}%
  \BibitemOpen
  \bibfield{author}{%
  \bibinfo {author} {\bibfnamefont{K.~L.}\ \bibnamefont{McDonald}}\ and\
  \bibinfo {author} {\bibfnamefont{D.~E.}\ \bibnamefont{Morrissey}},\ }%
  \bibfield{journal}{%
  \Doi{10.1007/JHEP02(2011)087}{\bibinfo {journal} {JHEP}}\ }%
  \textbf{\bibinfo {volume} {1102}},\ \bibinfo {pages} {087} (\bibinfo {year}
  {2011}),\ \Eprint{http://arxiv.org/abs/1010.5999}{arXiv:1010.5999 [hep-ph]}%
  \bibAnnoteFile{NoStop}{McDonald:2010fe}%
\bibitem{Williams:2011qb}%
  \BibitemOpen
  \bibfield{author}{%
  \bibinfo {author} {\bibfnamefont{M.}~\bibnamefont{Williams}}, \bibinfo
  {author} {\bibfnamefont{C.}~\bibnamefont{Burgess}}, \bibinfo {author}
  {\bibfnamefont{A.}~\bibnamefont{Maharana}},\ and\ \bibinfo {author}
  {\bibfnamefont{F.}~\bibnamefont{Quevedo}},\ }%
  \bibfield{journal}{%
  \Doi{10.1007/JHEP08(2011)106}{\bibinfo {journal} {JHEP}}\ }%
  \textbf{\bibinfo {volume} {1108}},\ \bibinfo {pages} {106} (\bibinfo {year}
  {2011}),\ \Eprint{http://arxiv.org/abs/1103.4556}{arXiv:1103.4556 [hep-ph]}%
  \bibAnnoteFile{NoStop}{Williams:2011qb}%
\bibitem{Abrahamyan:2011gv}%
  \BibitemOpen
  \bibfield{author}{%
  \bibinfo {author} {\bibfnamefont{S.}~\bibnamefont{Abrahamyan}} \emph{et~al.}
  (\bibinfo {collaboration} {APEX Collaboration}),\ }%
  \bibfield{journal}{%
  \Doi{10.1103/PhysRevLett.107.191804}{\bibinfo {journal} {Phys.Rev.Lett.}}\ }%
  \textbf{\bibinfo {volume} {107}},\ \bibinfo {pages} {191804} (\bibinfo {year}
  {2011}),\ \Eprint{http://arxiv.org/abs/1108.2750}{arXiv:1108.2750 [hep-ex]}%
  \bibAnnoteFile{NoStop}{Abrahamyan:2011gv}%
\bibitem{Archilli:2011zc}%
  \BibitemOpen
  \bibfield{author}{%
  \bibinfo {author} {\bibfnamefont{F.}~\bibnamefont{Archilli}}, \bibinfo
  {author} {\bibfnamefont{D.}~\bibnamefont{Babusci}}, \bibinfo {author}
  {\bibfnamefont{D.}~\bibnamefont{Badoni}}, \bibinfo {author}
  {\bibfnamefont{I.}~\bibnamefont{Balwierz}}, \bibinfo {author}
  {\bibfnamefont{G.}~\bibnamefont{Bencivenni}}, \emph{et~al.},\ }%
  \bibfield{journal}{%
  \Doi{10.1016/j.physletb.2011.11.033}{\bibinfo {journal} {Phys.Lett.}}\ }%
  \textbf{\bibinfo {volume} {B706}},\ \bibinfo {pages} {251} (\bibinfo {year}
  {2012}),\ \Eprint{http://arxiv.org/abs/1110.0411}{arXiv:1110.0411 [hep-ex]}%
  \bibAnnoteFile{NoStop}{Archilli:2011zc}%
\bibitem{Lees:2012ra}%
  \BibitemOpen
  \bibfield{author}{%
  \bibinfo {author} {\bibfnamefont{J.}~\bibnamefont{Lees}} \emph{et~al.}
  (\bibinfo {collaboration} {BaBar Collaboration}),\ }%
  \bibfield{journal}{%
  \Doi{10.1103/PhysRevLett.108.211801}{\bibinfo {journal} {Phys.Rev.Lett.}}\ }%
  \textbf{\bibinfo {volume} {108}},\ \bibinfo {pages} {211801} (\bibinfo {year}
  {2012}),\ \Eprint{http://arxiv.org/abs/1202.1313}{arXiv:1202.1313 [hep-ex]}%
  \bibAnnoteFile{NoStop}{Lees:2012ra}%
\bibitem{Davoudiasl:2012ag}%
  \BibitemOpen
  \bibfield{author}{%
  \bibinfo {author} {\bibfnamefont{H.}~\bibnamefont{Davoudiasl}}, \bibinfo
  {author} {\bibfnamefont{H.-S.}\ \bibnamefont{Lee}},\ and\ \bibinfo {author}
  {\bibfnamefont{W.~J.}\ \bibnamefont{Marciano}},\ }%
  \bibfield{journal}{%
  \Doi{10.1103/PhysRevD.85.115019}{\bibinfo {journal} {Phys.Rev.}}\ }%
  \textbf{\bibinfo {volume} {D85}},\ \bibinfo {pages} {115019} (\bibinfo {year}
  {2012}),\ \Eprint{http://arxiv.org/abs/1203.2947}{arXiv:1203.2947 [hep-ph]}%
  \bibAnnoteFile{NoStop}{Davoudiasl:2012ag}%
\bibitem{Kahn:2012br}%
  \BibitemOpen
  \bibfield{author}{%
  \bibinfo {author} {\bibfnamefont{Y.}~\bibnamefont{Kahn}}\ and\ \bibinfo
  {author} {\bibfnamefont{J.}~\bibnamefont{Thaler}},\ }%
  \bibfield{journal}{%
  \Doi{10.1103/PhysRevD.86.115012}{\bibinfo {journal} {Phys.Rev.}}\ }%
  \textbf{\bibinfo {volume} {D86}},\ \bibinfo {pages} {115012} (\bibinfo {year}
  {2012}),\ \Eprint{http://arxiv.org/abs/1209.0777}{arXiv:1209.0777 [hep-ph]}%
  \bibAnnoteFile{NoStop}{Kahn:2012br}%
\bibitem{Andreas:2012mt}%
  \BibitemOpen
  \bibfield{author}{%
  \bibinfo {author} {\bibfnamefont{S.}~\bibnamefont{Andreas}}, \bibinfo
  {author} {\bibfnamefont{C.}~\bibnamefont{Niebuhr}},\ and\ \bibinfo {author}
  {\bibfnamefont{A.}~\bibnamefont{Ringwald}},\ }%
  \bibfield{journal}{%
  \Doi{10.1103/PhysRevD.86.095019}{\bibinfo {journal} {Phys.Rev.}}\ }%
  \textbf{\bibinfo {volume} {D86}},\ \bibinfo {pages} {095019} (\bibinfo {year}
  {2012}),\ \Eprint{http://arxiv.org/abs/1209.6083}{arXiv:1209.6083 [hep-ph]}%
  \bibAnnoteFile{NoStop}{Andreas:2012mt}%
\bibitem{Essig:2013vha}%
  \BibitemOpen
  \bibfield{author}{%
  \bibinfo {author} {\bibfnamefont{R.}~\bibnamefont{Essig}}, \bibinfo {author}
  {\bibfnamefont{J.}~\bibnamefont{Mardon}}, \bibinfo {author}
  {\bibfnamefont{M.}~\bibnamefont{Papucci}}, \bibinfo {author}
  {\bibfnamefont{T.}~\bibnamefont{Volansky}},\ and\ \bibinfo {author}
  {\bibfnamefont{Y.-M.}\ \bibnamefont{Zhong}},\ }%
  \bibfield{journal}{%
  \Doi{10.1007/JHEP11(2013)167}{\bibinfo {journal} {JHEP}}\ }%
  \textbf{\bibinfo {volume} {1311}},\ \bibinfo {pages} {167} (\bibinfo {year}
  {2013}),\ \Eprint{http://arxiv.org/abs/1309.5084}{arXiv:1309.5084 [hep-ph]}%
  \bibAnnoteFile{NoStop}{Essig:2013vha}%
\bibitem{Davoudiasl:2013jma}%
  \BibitemOpen
  \bibfield{author}{%
  \bibinfo {author} {\bibfnamefont{H.}~\bibnamefont{Davoudiasl}}\ and\ \bibinfo
  {author} {\bibfnamefont{I.~M.}\ \bibnamefont{Lewis}},\ }%
  \bibfield{journal}{%
  \Doi{10.1103/PhysRevD.89.055026}{\bibinfo {journal} {Phys.Rev.}}\ }%
  \textbf{\bibinfo {volume} {D89}},\ \bibinfo {pages} {055026} (\bibinfo {year}
  {2014}),\ \Eprint{http://arxiv.org/abs/1309.6640}{arXiv:1309.6640 [hep-ph]}%
  \bibAnnoteFile{NoStop}{Davoudiasl:2013jma}%
\bibitem{Morrissey:2014yma}%
  \BibitemOpen
  \bibfield{author}{%
  \bibinfo {author} {\bibfnamefont{D.~E.}\ \bibnamefont{Morrissey}}\ and\
  \bibinfo {author} {\bibfnamefont{A.~P.}\ \bibnamefont{Spray}}}%
   (\bibinfo {year} {2014}),\
  \Eprint{http://arxiv.org/abs/1402.4817}{arXiv:1402.4817 [hep-ph]}%
  \bibAnnoteFile{NoStop}{Morrissey:2014yma}%
\bibitem{Babusci:2014sta}%
  \BibitemOpen
  \bibfield{author}{%
  \bibinfo {author} {\bibfnamefont{D.}~\bibnamefont{Babusci}} \emph{et~al.}
  (\bibinfo {collaboration} {KLOE-2 Collaboration})}%
   (\bibinfo {year} {2014}),\
  \Eprint{http://arxiv.org/abs/1404.7772}{arXiv:1404.7772 [hep-ex]}%
  \bibAnnoteFile{NoStop}{Babusci:2014sta}%
\bibitem{Izaguirre:2015yja}%
  \BibitemOpen
  \bibfield{author}{%
  \bibinfo {author} {\bibfnamefont{E.}~\bibnamefont{Izaguirre}}, \bibinfo
  {author} {\bibfnamefont{G.}~\bibnamefont{Krnjaic}}, \bibinfo {author}
  {\bibfnamefont{P.}~\bibnamefont{Schuster}},\ and\ \bibinfo {author}
  {\bibfnamefont{N.}~\bibnamefont{Toro}}}%
   (\bibinfo {year} {2015}),\
  \Eprint{http://arxiv.org/abs/1505.00011}{arXiv:1505.00011 [hep-ph]}%
  \bibAnnoteFile{NoStop}{Izaguirre:2015yja}%
\bibitem{Feng:2017drg}%
  \BibitemOpen
  \bibfield{author}{%
  \bibinfo {author} {\bibfnamefont{J.~L.}\ \bibnamefont{Feng}}\ and\ \bibinfo
  {author} {\bibfnamefont{J.}~\bibnamefont{Smolinsky}},\ }%
  \bibfield{journal}{%
  \Doi{10.1103/PhysRevD.96.095022}{\bibinfo {journal} {Phys. Rev.}}\ }%
  \textbf{\bibinfo {volume} {D96}},\ \bibinfo {pages} {095022} (\bibinfo {year}
  {2017}),\ \Eprint{http://arxiv.org/abs/1707.03835}{arXiv:1707.03835
  [hep-ph]}%
  \bibAnnoteFile{NoStop}{Feng:2017drg}%
\bibitem{Ade:2015xua}%
  \BibitemOpen
  \bibfield{author}{%
  \bibinfo {author} {\bibfnamefont{P.~A.~R.}\ \bibnamefont{Ade}} \emph{et~al.}
  (\bibinfo {collaboration} {Planck}),\ }%
  \bibfield{journal}{%
  \Doi{10.1051/0004-6361/201525830}{\bibinfo {journal} {Astron. Astrophys.}}\
  }%
  \textbf{\bibinfo {volume} {594}},\ \bibinfo {pages} {A13} (\bibinfo {year}
  {2016}),\ \Eprint{http://arxiv.org/abs/1502.01589}{arXiv:1502.01589
  [astro-ph.CO]}%
  \bibAnnoteFile{NoStop}{Ade:2015xua}%
\bibitem{Aghanim:2018eyx}%
  \BibitemOpen
  \bibfield{author}{%
  \bibinfo {author} {\bibfnamefont{N.}~\bibnamefont{Aghanim}} \emph{et~al.}
  (\bibinfo {collaboration} {Planck})}%
   (\bibinfo {year} {2018}),\
  \Eprint{http://arxiv.org/abs/1807.06209}{arXiv:1807.06209 [astro-ph.CO]}%
  \bibAnnoteFile{NoStop}{Aghanim:2018eyx}%
\bibitem{Krnjaic:2015mbs}%
  \BibitemOpen
  \bibfield{author}{%
  \bibinfo {author} {\bibfnamefont{G.}~\bibnamefont{Krnjaic}},\ }%
  \bibfield{journal}{%
  \Doi{10.1103/PhysRevD.94.073009}{\bibinfo {journal} {Phys. Rev.}}\ }%
  \textbf{\bibinfo {volume} {D94}},\ \bibinfo {pages} {073009} (\bibinfo {year}
  {2016}),\ \Eprint{http://arxiv.org/abs/1512.04119}{arXiv:1512.04119
  [hep-ph]}%
  \bibAnnoteFile{NoStop}{Krnjaic:2015mbs}%
\bibitem{Griest:1990kh}%
  \BibitemOpen
  \bibfield{author}{%
  \bibinfo {author} {\bibfnamefont{K.}~\bibnamefont{Griest}}\ and\ \bibinfo
  {author} {\bibfnamefont{D.}~\bibnamefont{Seckel}},\ }%
  \bibfield{journal}{%
  \Doi{10.1103/PhysRevD.43.3191}{\bibinfo {journal} {Phys. Rev.}}\ }%
  \textbf{\bibinfo {volume} {D43}},\ \bibinfo {pages} {3191} (\bibinfo {year}
  {1991})%
  \bibAnnoteFile{NoStop}{Griest:1990kh}%
\bibitem{Pospelov:2007mp}%
  \BibitemOpen
  \bibfield{author}{%
  \bibinfo {author} {\bibfnamefont{M.}~\bibnamefont{Pospelov}}, \bibinfo
  {author} {\bibfnamefont{A.}~\bibnamefont{Ritz}},\ and\ \bibinfo {author}
  {\bibfnamefont{M.~B.}\ \bibnamefont{Voloshin}},\ }%
  \bibfield{journal}{%
  \Doi{10.1016/j.physletb.2008.02.052}{\bibinfo {journal} {Phys.Lett.}}\ }%
  \textbf{\bibinfo {volume} {B662}},\ \bibinfo {pages} {53} (\bibinfo {year}
  {2008}),\ \Eprint{http://arxiv.org/abs/0711.4866}{arXiv:0711.4866 [hep-ph]}%
  \bibAnnoteFile{NoStop}{Pospelov:2007mp}%
\bibitem{DAgnolo:2015ujb}%
  \BibitemOpen
  \bibfield{author}{%
  \bibinfo {author} {\bibfnamefont{R.~T.}\ \bibnamefont{D'Agnolo}}\ and\
  \bibinfo {author} {\bibfnamefont{J.~T.}\ \bibnamefont{Ruderman}},\ }%
  \bibfield{journal}{%
  \Doi{10.1103/PhysRevLett.115.061301}{\bibinfo {journal} {Phys. Rev. Lett.}}\
  }%
  \textbf{\bibinfo {volume} {115}},\ \bibinfo {pages} {061301} (\bibinfo {year}
  {2015}),\ \Eprint{http://arxiv.org/abs/1505.07107}{arXiv:1505.07107
  [hep-ph]}%
  \bibAnnoteFile{NoStop}{DAgnolo:2015ujb}%
\bibitem{Cline:2017tka}%
  \BibitemOpen
  \bibfield{author}{%
  \bibinfo {author} {\bibfnamefont{J.~M.}\ \bibnamefont{Cline}}, \bibinfo
  {author} {\bibfnamefont{H.}~\bibnamefont{Liu}}, \bibinfo {author}
  {\bibfnamefont{T.}~\bibnamefont{Slatyer}},\ and\ \bibinfo {author}
  {\bibfnamefont{W.}~\bibnamefont{Xue}},\ }%
  \bibfield{journal}{%
  \Doi{10.1103/PhysRevD.96.083521}{\bibinfo {journal} {Phys. Rev.}}\ }%
  \textbf{\bibinfo {volume} {D96}},\ \bibinfo {pages} {083521} (\bibinfo {year}
  {2017}),\ \Eprint{http://arxiv.org/abs/1702.07716}{arXiv:1702.07716
  [hep-ph]}%
  \bibAnnoteFile{NoStop}{Cline:2017tka}%
\bibitem{Akesson:2018vlm}%
  \BibitemOpen
  \bibfield{author}{%
  \bibinfo {author} {\bibfnamefont{T.}~\bibnamefont{Akesson}} \emph{et~al.}
  (\bibinfo {collaboration} {LDMX})}%
   (\bibinfo {year} {2018}),\
  \Eprint{http://arxiv.org/abs/1808.05219}{arXiv:1808.05219 [hep-ex]}%
  \bibAnnoteFile{NoStop}{Akesson:2018vlm}%
\bibitem{Battaglieri:2016ggd}%
  \BibitemOpen
  \bibfield{author}{%
  \bibinfo {author} {\bibfnamefont{M.}~\bibnamefont{Battaglieri}} \emph{et~al.}
  (\bibinfo {collaboration} {BDX})}%
   (\bibinfo {year} {2016}),\
  \Eprint{http://arxiv.org/abs/1607.01390}{arXiv:1607.01390 [hep-ex]}%
  \bibAnnoteFile{NoStop}{Battaglieri:2016ggd}%
\bibitem{deNiverville:2018dbu}%
  \BibitemOpen
  \bibfield{author}{%
  \bibinfo {author} {\bibfnamefont{P.}~\bibnamefont{deNiverville}}\ and\
  \bibinfo {author} {\bibfnamefont{C.}~\bibnamefont{Frugiuele}},\ }%
  \bibfield{journal}{%
  \Doi{10.1103/PhysRevD.99.051701}{\bibinfo {journal} {Phys. Rev.}}\ }%
  \textbf{\bibinfo {volume} {D99}},\ \bibinfo {pages} {051701} (\bibinfo {year}
  {2019}),\ \Eprint{http://arxiv.org/abs/1807.06501}{arXiv:1807.06501
  [hep-ph]}%
  \bibAnnoteFile{NoStop}{deNiverville:2018dbu}%
\bibitem{Akimov:2019rhz}%
  \BibitemOpen
  \bibfield{author}{%
  \bibinfo {author} {\bibfnamefont{D.}~\bibnamefont{Akimov}} \emph{et~al.}
  (\bibinfo {collaboration} {COHERENT})}%
   (\bibinfo {year} {2019}),\
  \Eprint{http://arxiv.org/abs/1909.05913}{arXiv:1909.05913 [hep-ex]}%
  \bibAnnoteFile{NoStop}{Akimov:2019rhz}%
\bibitem{Akimov:2019xdj}%
  \BibitemOpen
  \bibfield{author}{%
  \bibinfo {author} {\bibfnamefont{D.}~\bibnamefont{Akimov}} \emph{et~al.}
  (\bibinfo {collaboration} {COHERENT})}%
   (\bibinfo {year} {2019}),\
  \Eprint{http://arxiv.org/abs/1911.06422}{arXiv:1911.06422 [hep-ex]}%
  \bibAnnoteFile{NoStop}{Akimov:2019xdj}%
\bibitem{CCM}%
  \BibitemOpen
  \bibfield{author}{%
  \bibinfo {author} {\bibfnamefont{R.}~\bibnamefont{Van~de Water}},\ }%
  \enquote{\bibinfo {title} {{Searching for Sterile Neutrinos with the Coherent
  CAPTAIN-Mills Detector at the Los Alamos Neutron Science Center}},}\
  \Eprint{http://arxiv.org/abs/APS April Meeting 2019}{APS April Meeting
  2019},\ \url{http://meetings.aps.org/Meeting/APR19/Session/Z14.9}%
  \bibAnnoteFile{NoStop}{CCM}%
\bibitem{Aubert:2008as}%
  \BibitemOpen
  \bibfield{author}{%
  \bibinfo {author} {\bibfnamefont{B.}~\bibnamefont{Aubert}} \emph{et~al.}
  (\bibinfo {collaboration} {BaBar}),\ }%
  in\ \emph{\bibinfo {booktitle} {{Proceedings, 34th International Conference
  on High Energy Physics (ICHEP 2008): Philadelphia, Pennsylvania, July
  30-August 5, 2008}}}\ (\bibinfo {year} {2008})\
  \Eprint{http://arxiv.org/abs/0808.0017}{arXiv:0808.0017 [hep-ex]},\
  \url{http://www-public.slac.stanford.edu/sciDoc/docMeta.aspx?slacPubNumber=s%
lac-pub-13328}%
  \bibAnnoteFile{NoStop}{Aubert:2008as}%
\bibitem{Hook:2010tw}%
  \BibitemOpen
  \bibfield{author}{%
  \bibinfo {author} {\bibfnamefont{A.}~\bibnamefont{Hook}}, \bibinfo {author}
  {\bibfnamefont{E.}~\bibnamefont{Izaguirre}},\ and\ \bibinfo {author}
  {\bibfnamefont{J.~G.}\ \bibnamefont{Wacker}},\ }%
  \bibfield{journal}{%
  \Doi{10.1155/2011/859762}{\bibinfo {journal} {Adv. High Energy Phys.}}\ }%
  \textbf{\bibinfo {volume} {2011}},\ \bibinfo {pages} {859762} (\bibinfo
  {year} {2011}),\ \Eprint{http://arxiv.org/abs/1006.0973}{arXiv:1006.0973
  [hep-ph]}%
  \bibAnnoteFile{NoStop}{Hook:2010tw}%
\bibitem{Curtin:2014cca}%
  \BibitemOpen
  \bibfield{author}{%
  \bibinfo {author} {\bibfnamefont{D.}~\bibnamefont{Curtin}}, \bibinfo {author}
  {\bibfnamefont{R.}~\bibnamefont{Essig}}, \bibinfo {author}
  {\bibfnamefont{S.}~\bibnamefont{Gori}},\ and\ \bibinfo {author}
  {\bibfnamefont{J.}~\bibnamefont{Shelton}},\ }%
  \bibfield{journal}{%
  \Doi{10.1007/JHEP02(2015)157}{\bibinfo {journal} {JHEP}}\ }%
  \textbf{\bibinfo {volume} {02}},\ \bibinfo {pages} {157} (\bibinfo {year}
  {2015}),\ \Eprint{http://arxiv.org/abs/1412.0018}{arXiv:1412.0018 [hep-ph]}%
  \bibAnnoteFile{NoStop}{Curtin:2014cca}%
\bibitem{Battaglieri:2017aum}%
  \BibitemOpen
  \bibfield{author}{%
  \bibinfo {author} {\bibfnamefont{M.}~\bibnamefont{Battaglieri}}
  \emph{et~al.},\ }%
  in\ \emph{\bibinfo {booktitle} {{U.S. Cosmic Visions: New Ideas in Dark
  Matter College Park, MD, USA, March 23-25, 2017}}}\ (\bibinfo {year} {2017})\
  \Eprint{http://arxiv.org/abs/1707.04591}{arXiv:1707.04591 [hep-ph]},\
  \url{http://lss.fnal.gov/archive/2017/conf/fermilab-conf-17-282-ae-ppd-t.pdf%
}%
  \bibAnnoteFile{NoStop}{Battaglieri:2017aum}%
\bibitem{mg}%
  \BibitemOpen
  \bibfield{author}{%
  \bibinfo {author} {\bibfnamefont{J.}~\bibnamefont{Alwall}}, \bibinfo {author}
  {\bibfnamefont{R.}~\bibnamefont{Frederix}}, \bibinfo {author}
  {\bibfnamefont{S.}~\bibnamefont{Frixione}}, \bibinfo {author}
  {\bibfnamefont{V.}~\bibnamefont{Hirschi}}, \bibinfo {author}
  {\bibfnamefont{F.}~\bibnamefont{Maltoni}}, \bibinfo {author}
  {\bibfnamefont{O.}~\bibnamefont{Mattelaer}}, \bibinfo {author}
  {\bibfnamefont{H.~S.}\ \bibnamefont{Shao}}, \bibinfo {author}
  {\bibfnamefont{T.}~\bibnamefont{Stelzer}}, \bibinfo {author}
  {\bibfnamefont{P.}~\bibnamefont{Torrielli}},\ and\ \bibinfo {author}
  {\bibfnamefont{M.}~\bibnamefont{Zaro}},\ }%
  \bibfield{journal}{%
  \Doi{10.1007/JHEP07(2014)079}{\bibinfo {journal} {JHEP}}\ }%
  \textbf{\bibinfo {volume} {07}},\ \bibinfo {pages} {079} (\bibinfo {year}
  {2014}),\ \Eprint{http://arxiv.org/abs/1405.0301}{arXiv:1405.0301 [hep-ph]}%
  \bibAnnoteFile{NoStop}{mg}%
\bibitem{Dharmapalan:2012xp}%
  \BibitemOpen
  \bibfield{author}{%
  \bibinfo {author} {\bibfnamefont{R.}~\bibnamefont{Dharmapalan}} \emph{et~al.}
  (\bibinfo {collaboration} {MiniBooNE Collaboration})}%
   (\bibinfo {year} {2012}),\
  \Eprint{http://arxiv.org/abs/1211.2258}{arXiv:1211.2258 [hep-ex]}%
  \bibAnnoteFile{NoStop}{Dharmapalan:2012xp}%
\bibitem{Batell:2014yra}%
  \BibitemOpen
  \bibfield{author}{%
  \bibinfo {author} {\bibfnamefont{B.}~\bibnamefont{Batell}}, \bibinfo {author}
  {\bibfnamefont{P.}~\bibnamefont{deNiverville}}, \bibinfo {author}
  {\bibfnamefont{D.}~\bibnamefont{McKeen}}, \bibinfo {author}
  {\bibfnamefont{M.}~\bibnamefont{Pospelov}},\ and\ \bibinfo {author}
  {\bibfnamefont{A.}~\bibnamefont{Ritz}}}%
   (\bibinfo {year} {2014}),\
  \Eprint{http://arxiv.org/abs/1405.7049}{arXiv:1405.7049 [hep-ph]}%
  \bibAnnoteFile{NoStop}{Batell:2014yra}%
\bibitem{Gorbunov:2014wqa}%
  \BibitemOpen
  \bibfield{author}{%
  \bibinfo {author} {\bibfnamefont{D.}~\bibnamefont{Gorbunov}}, \bibinfo
  {author} {\bibfnamefont{A.}~\bibnamefont{Makarov}},\ and\ \bibinfo {author}
  {\bibfnamefont{I.}~\bibnamefont{Timiryasov}},\ }%
  \bibfield{journal}{%
  \Doi{10.1103/PhysRevD.91.035027}{\bibinfo {journal} {Phys. Rev.}}\ }%
  \textbf{\bibinfo {volume} {D91}},\ \bibinfo {pages} {035027} (\bibinfo {year}
  {2015}),\ \Eprint{http://arxiv.org/abs/1411.4007}{arXiv:1411.4007 [hep-ph]}%
  \bibAnnoteFile{NoStop}{Gorbunov:2014wqa}%
\bibitem{Blumlein:2013cua}%
  \BibitemOpen
  \bibfield{author}{%
  \bibinfo {author} {\bibfnamefont{J.}~\bibnamefont{Blümlein}}\ and\ \bibinfo
  {author} {\bibfnamefont{J.}~\bibnamefont{Brunner}},\ }%
  \bibfield{journal}{%
  \Doi{10.1016/j.physletb.2014.02.029}{\bibinfo {journal} {Phys. Lett.}}\ }%
  \textbf{\bibinfo {volume} {B731}},\ \bibinfo {pages} {320} (\bibinfo {year}
  {2014}),\ \Eprint{http://arxiv.org/abs/1311.3870}{arXiv:1311.3870 [hep-ph]}%
  \bibAnnoteFile{NoStop}{Blumlein:2013cua}%
\bibitem{AguilarArevalo:2008yp}%
  \BibitemOpen
  \bibfield{author}{%
  \bibinfo {author} {\bibfnamefont{A.}~\bibnamefont{Aguilar-Arevalo}}
  \emph{et~al.} (\bibinfo {collaboration} {MiniBooNE Collaboration}),\ }%
  \bibfield{journal}{%
  \Doi{10.1103/PhysRevD.79.072002}{\bibinfo {journal} {Phys.Rev.}}\ }%
  \textbf{\bibinfo {volume} {D79}},\ \bibinfo {pages} {072002} (\bibinfo {year}
  {2009}),\ \Eprint{http://arxiv.org/abs/0806.1449}{arXiv:0806.1449 [hep-ex]}%
  \bibAnnoteFile{NoStop}{AguilarArevalo:2008yp}%
\bibitem{Bonesini:2001iz}%
  \BibitemOpen
  \bibfield{author}{%
  \bibinfo {author} {\bibfnamefont{M.}~\bibnamefont{Bonesini}}, \bibinfo
  {author} {\bibfnamefont{A.}~\bibnamefont{Marchionni}}, \bibinfo {author}
  {\bibfnamefont{F.}~\bibnamefont{Pietropaolo}},\ and\ \bibinfo {author}
  {\bibfnamefont{T.}~\bibnamefont{Tabarelli~de Fatis}},\ }%
  \bibfield{journal}{%
  \Doi{10.1007/s100520100656}{\bibinfo {journal} {Eur. Phys. J.}}\ }%
  \textbf{\bibinfo {volume} {C20}},\ \bibinfo {pages} {13} (\bibinfo {year}
  {2001}),\ \Eprint{http://arxiv.org/abs/hep-ph/0101163}{arXiv:hep-ph/0101163
  [hep-ph]}%
  \bibAnnoteFile{NoStop}{Bonesini:2001iz}%
\bibitem{Burman:1989ds}%
  \BibitemOpen
  \bibfield{author}{%
  \bibinfo {author} {\bibfnamefont{R.}~\bibnamefont{Burman}}\ and\ \bibinfo
  {author} {\bibfnamefont{E.}~\bibnamefont{Smith}},\ }%
  \bibfield{journal}{%
  \bibinfo {journal} {LA-11502-MS, DE-98-011120, UC-414, Los Alamos}}%
   (\bibinfo {year} {1989})%
  \bibAnnoteFile{NoStop}{Burman:1989ds}%
\bibitem{Aguilar-Arevalo:2018wea}%
  \BibitemOpen
  \bibfield{author}{%
  \bibinfo {author} {\bibfnamefont{A.~A.}\ \bibnamefont{Aguilar-Arevalo}}
  \emph{et~al.} (\bibinfo {collaboration} {MiniBooNE DM}),\ }%
  \bibfield{journal}{%
  \Doi{10.1103/PhysRevD.98.112004}{\bibinfo {journal} {Phys. Rev.}}\ }%
  \textbf{\bibinfo {volume} {D98}},\ \bibinfo {pages} {112004} (\bibinfo {year}
  {2018}),\ \Eprint{http://arxiv.org/abs/1807.06137}{arXiv:1807.06137
  [hep-ex]}%
  \bibAnnoteFile{NoStop}{Aguilar-Arevalo:2018wea}%
\bibitem{Shoemaker:2011vi}%
  \BibitemOpen
  \bibfield{author}{%
  \bibinfo {author} {\bibfnamefont{I.~M.}\ \bibnamefont{Shoemaker}}\ and\
  \bibinfo {author} {\bibfnamefont{L.}~\bibnamefont{Vecchi}},\ }%
  \bibfield{journal}{%
  \Doi{10.1103/PhysRevD.86.015023}{\bibinfo {journal} {Phys. Rev.}}\ }%
  \textbf{\bibinfo {volume} {D86}},\ \bibinfo {pages} {015023} (\bibinfo {year}
  {2012}),\ \Eprint{http://arxiv.org/abs/1112.5457}{arXiv:1112.5457 [hep-ph]}%
  \bibAnnoteFile{NoStop}{Shoemaker:2011vi}%
\bibitem{Dror:2018wfl}%
  \BibitemOpen
  \bibfield{author}{%
  \bibinfo {author} {\bibfnamefont{J.~A.}\ \bibnamefont{Dror}}, \bibinfo
  {author} {\bibfnamefont{R.}~\bibnamefont{Lasenby}},\ and\ \bibinfo {author}
  {\bibfnamefont{M.}~\bibnamefont{Pospelov}},\ }%
  \bibfield{journal}{%
  \Doi{10.1103/PhysRevD.99.055016}{\bibinfo {journal} {Phys. Rev.}}\ }%
  \textbf{\bibinfo {volume} {D99}},\ \bibinfo {pages} {055016} (\bibinfo {year}
  {2019}),\ \Eprint{http://arxiv.org/abs/1811.00595}{arXiv:1811.00595
  [hep-ph]}%
  \bibAnnoteFile{NoStop}{Dror:2018wfl}%
\bibitem{Dror:2017ehi}%
  \BibitemOpen
  \bibfield{author}{%
  \bibinfo {author} {\bibfnamefont{J.~A.}\ \bibnamefont{Dror}}, \bibinfo
  {author} {\bibfnamefont{R.}~\bibnamefont{Lasenby}},\ and\ \bibinfo {author}
  {\bibfnamefont{M.}~\bibnamefont{Pospelov}},\ }%
  \bibfield{journal}{%
  \Doi{10.1103/PhysRevLett.119.141803}{\bibinfo {journal} {Phys. Rev. Lett.}}\
  }%
  \textbf{\bibinfo {volume} {119}},\ \bibinfo {pages} {141803} (\bibinfo {year}
  {2017}),\ \Eprint{http://arxiv.org/abs/1705.06726}{arXiv:1705.06726
  [hep-ph]}%
  \bibAnnoteFile{NoStop}{Dror:2017ehi}%
\bibitem{Dror:2017nsg}%
  \BibitemOpen
  \bibfield{author}{%
  \bibinfo {author} {\bibfnamefont{J.~A.}\ \bibnamefont{Dror}}, \bibinfo
  {author} {\bibfnamefont{R.}~\bibnamefont{Lasenby}},\ and\ \bibinfo {author}
  {\bibfnamefont{M.}~\bibnamefont{Pospelov}},\ }%
  \bibfield{journal}{%
  \Doi{10.1103/PhysRevD.96.075036}{\bibinfo {journal} {Phys. Rev.}}\ }%
  \textbf{\bibinfo {volume} {D96}},\ \bibinfo {pages} {075036} (\bibinfo {year}
  {2017}),\ \Eprint{http://arxiv.org/abs/1707.01503}{arXiv:1707.01503
  [hep-ph]}%
  \bibAnnoteFile{NoStop}{Dror:2017nsg}%
\bibitem{Angle:2011th}%
  \BibitemOpen
  \bibfield{author}{%
  \bibinfo {author} {\bibfnamefont{J.}~\bibnamefont{Angle}} \emph{et~al.}
  (\bibinfo {collaboration} {XENON10}),\ }%
  \bibfield{journal}{%
  \Doi{10.1103/PhysRevLett.110.249901, 10.1103/PhysRevLett.107.051301}{\bibinfo
  {journal} {Phys. Rev. Lett.}}\ }%
  \textbf{\bibinfo {volume} {107}},\ \bibinfo {pages} {051301} (\bibinfo {year}
  {2011}),\ \bibinfo {note} {[Erratum: Phys. Rev. Lett.110,249901(2013)]},\
  \Eprint{http://arxiv.org/abs/1104.3088}{arXiv:1104.3088 [astro-ph.CO]}%
  \bibAnnoteFile{NoStop}{Angle:2011th}%
\bibitem{Essig:2017kqs}%
  \BibitemOpen
  \bibfield{author}{%
  \bibinfo {author} {\bibfnamefont{R.}~\bibnamefont{Essig}}, \bibinfo {author}
  {\bibfnamefont{T.}~\bibnamefont{Volansky}},\ and\ \bibinfo {author}
  {\bibfnamefont{T.-T.}\ \bibnamefont{Yu}},\ }%
  \bibfield{journal}{%
  \Doi{10.1103/PhysRevD.96.043017}{\bibinfo {journal} {Phys. Rev.}}\ }%
  \textbf{\bibinfo {volume} {D96}},\ \bibinfo {pages} {043017} (\bibinfo {year}
  {2017}),\ \Eprint{http://arxiv.org/abs/1703.00910}{arXiv:1703.00910
  [hep-ph]}%
  \bibAnnoteFile{NoStop}{Essig:2017kqs}%
\bibitem{Hall:2009bx}%
  \BibitemOpen
  \bibfield{author}{%
  \bibinfo {author} {\bibfnamefont{L.~J.}\ \bibnamefont{Hall}}, \bibinfo
  {author} {\bibfnamefont{K.}~\bibnamefont{Jedamzik}}, \bibinfo {author}
  {\bibfnamefont{J.}~\bibnamefont{March-Russell}},\ and\ \bibinfo {author}
  {\bibfnamefont{S.~M.}\ \bibnamefont{West}},\ }%
  \bibfield{journal}{%
  \Doi{10.1007/JHEP03(2010)080}{\bibinfo {journal} {JHEP}}\ }%
  \textbf{\bibinfo {volume} {03}},\ \bibinfo {pages} {080} (\bibinfo {year}
  {2010}),\ \Eprint{http://arxiv.org/abs/0911.1120}{arXiv:0911.1120 [hep-ph]}%
  \bibAnnoteFile{NoStop}{Hall:2009bx}%
\end{thebibliography}%
\end{document}